\documentclass[11pt]{article}

\usepackage{verbatim}
\usepackage{amsmath}
\usepackage{amsfonts}
\usepackage{amssymb}
\usepackage{bbm}
\usepackage{bbold}
\usepackage{latexsym}
\usepackage{rotating} 
\usepackage{hhline} 
\usepackage{epsfig}
\usepackage[usenames,dvipsnames]{color}
\usepackage{graphicx}
\usepackage{enumitem}
\usepackage[normalem]{ulem} 
\usepackage[table]{xcolor}
\usepackage{subfig}
\usepackage{multirow}
\usepackage{mathrsfs}
\usepackage{blkarray}
\usepackage{amscd}
\usepackage{cite}
\usepackage{arydshln}
\usepackage{longtable} 
\usepackage{tikz}
\usetikzlibrary{arrows.meta} 

\usepackage{color}   
\usepackage{hyperref}
\hypersetup{
    colorlinks=true, 
    linktoc=all,     
    citecolor=black,
    filecolor=black,
    linkcolor=black    
}

\renewcommand{\d}{\mathrm{d}}

\DeclareMathSymbol{\mg}{\mathrel}{symbols}{"1D}

\newcommand{\bes}{\begin{split}}
\newcommand{\ees}{\end{split}}

\newcommand{\ga}{\alpha}
\newcommand{\gb}{\beta}
\renewcommand{\gg}{\gamma}
\newcommand{\gd}{\delta}
\renewcommand{\ge}{\epsilon}
\newcommand{\gve}{\varepsilon}
\newcommand{\gf}{\phi}
\newcommand{\gvf}{\varphi}

\newcommand{\gx}{\xi}
\newcommand{\gm}{\mu}
\newcommand{\gn}{\nu}

\newcommand{\gk}{\kappa}
\newcommand{\gl}{\lambda}
\newcommand{\gr}{\rho}
\newcommand{\gth}{\theta}

\newcommand{\gs}{\sigma}
\newcommand{\gt}{\tau}
\newcommand{\go}{\omega}
\newcommand{\gz}{\zeta}
\newcommand{\gp}{\pi}
\newcommand{\gps}{\psi}

\newcommand{\gch}{\chi}
\newcommand{\gG}{\Gamma}

\newcommand{\gF}{\Phi}

\newcommand{\gX}{\Xi}
\newcommand{\gL}{\Lambda}
\newcommand{\gS}{\Sigma}
\newcommand{\gTh}{\Theta}
\newcommand{\gO}{\Omega}

\newcommand{\gPs}{\Psi}

\newcommand{\gY}{\Upsilon}
\newcommand{\cA}{{\cal A}}

\newcommand{\cD}{{\cal D}}

\newcommand{\cF}{{\cal F}}

\newcommand{\cL}{{\cal L}}
\newcommand{\cM}{{\cal M}}
\newcommand{\cN}{{\cal N}}
\newcommand{\cO}{{\cal O}}
\newcommand{\cP}{{\cal P}}
\newcommand{\cQ}{{\cal Q}}
\newcommand{\cR}{{\cal R}}

\newcommand{\cV}{{\cal V}}

\newcommand{\ua}{{\underline a}}

\newcommand{\uc}{{\underline c}}

\newcommand{\um}{{\underline m}}

\newcommand{\tQ}{{\widetilde Q}}

\newcommand{\tZ}{{\widetilde Z}}
\newcommand{\uga}{{\underline \alpha}}
\newcommand{\ugb}{{\underline\beta}}
\newcommand{\ugg}{{\underline\gamma}}

\newcommand{\ugm}{{\underline\mu}}
\newcommand{\ugn}{{\underline\nu}}

\newcommand{\Tr}{\mbox{Tr}}
\newcommand{\tr}{\text{tr}}

\newcommand{\Id}{\mathbbm{1}}
\newcommand{\slashed}{\hspace{-1.1ex}/}
\newcommand{\Slashed}{\hspace{-1.4ex}/\hspace{.2ex}}
\newcommand{\lra}{\longrightarrow}

\newcommand{\ra}{\rightarrow}

\newcommand{\der}{\partial}

\newcommand{\beq}{\begin{equation}}
\newcommand{\eeq}{\end{equation}}
\newcommand{\barr}{\begin{array}}
\newcommand{\earr}{\end{array}}
\newcommand{\equ}[1]{\begin{gather} #1 \end{gather}}
\newcommand{\sequ}[2]{\begin{subequations}\label{#1}\begin{gather} #2 \end{gather}\end{subequations}}

\newcommand{\enums}[1]{\begin{enumerate} #1 \end{enumerate}}

\newcommand{\arry}[2]{\begin{array}{#1} #2 \end{array}}

\newcommand{\pmtrx}[1]{\begin{pmatrix} #1 \end{pmatrix}}

\newcommand{\sfrac}[2]{\mbox{$\frac{#1}{#2}$}}
\newcounter{oldcounter}

\newcommand{\bder}{\bar\partial}
\newcommand{\bff}{{\bar f}}

\newcommand{\bh}{{\bar h}}

\newcommand{\bk}{{\bar k}}

\newcommand{\bn}{{\bar n}}

\newcommand{\bs}{{\bar s}}

\newcommand{\bz}{{\bar z}}

\newcommand{\bD}{{\overline D}}

\newcommand{\bF}{{\overline F}}

\newcommand{\bK}{{\overline K}}

\newcommand{\bQ}{{\overline Q}}

\newcommand{\bU}{{\overline U}}

\newcommand{\bW}{{\overline W}}

\newcommand{\bge}{{\bar\epsilon}}

\newcommand{\bgf}{{\bar\phi}}
\newcommand{\bgvf}{{\bar\varphi}}

\newcommand{\bgl}{{\bar\lambda}}
\newcommand{\bgr}{{\bar\rho}}
\newcommand{\bgth}{{\bar\theta}}
\newcommand{\bgs}{{\bar\sigma}}

\newcommand{\bgo}{{\bar\omega}}
\newcommand{\bgz}{{\bar\zeta}}

\newcommand{\bgps}{{\bar\psi}}

\newcommand{\bgch}{{\bar\chi}}
\newcommand{\bgG}{{\overline\Gamma}}

\newcommand{\bgF}{{\overline\Phi}}

\newcommand{\bgX}{{\overline\Xi}}
\newcommand{\bgL}{{\overline\Lambda}}
\newcommand{\bgS}{{\overline\Sigma}}
\newcommand{\bgTh}{{\overline\Theta}}
\newcommand{\bgO}{{\overline\Omega}}

\newcommand{\bgPs}{{\overline\Psi}}

\newcommand{\bgY}{{\overline\Upsilon}}
\newcommand{\bcD}{{\overline{\cal D}}}

\newcommand{\BgL}{{\boldsymbol \Lambda}}

\newcommand{\BgF}{{\boldsymbol \Phi}}

\newcommand{\BbgF}{\boldsymbol{\overline\Phi}}

\newcommand{\BbgL}{\boldsymbol{\overline\Lambda}}

\newcommand{\tgg}{{\tilde \gamma}}

\newcommand{\Intr}{\mathbbm{Z}}
\newcommand{\Cplx}{\mathbbm{C}}
\newcommand{\Real}{\mathbbm{R}}

\newcommand{\ba}[2]{\[\begin{array}{#2}\label{#1}}
\newcommand{\ea}{\end{array}\]}
\newcommand{\be}{\begin{equation}}
\newcommand{\ee}{\end{equation}}
\newcommand{\bea}{\begin{eqnarray}}
\newcommand{\eea}{\end{eqnarray}}

\newcommand{\sm}{{\,\mbox{-}}}

\begin{document}

\thispagestyle{empty}

\begin{flushright}
LTH-1318\\ 
\end{flushright}
\vskip 1 cm
\begin{center}
{\Large {\bf 
The fate of discrete torsion on resolved heterotic 
\\[1ex]
 $\boldsymbol{\Intr_2\times\Intr_2}$ orbifolds  using (0,2) GLSMs
} 
}
\\[0pt]

\bigskip
\bigskip {\large
{\bf A.E.~Faraggi$^{a,}$}\footnote{
E-mail: alon.faraggi@liverpool.ac.uk},
{\bf S.~Groot Nibbelink$^{b,}$}\footnote{
E-mail: s.groot.nibbelink@hr.nl},
{\bf   M.~Hurtado Heredia$^{a,}$}\footnote{
E-mail: martin.hurtado@liv.ac.uk}
\bigskip }\\[0pt]
\vspace{0.23cm}
${}^a$ {\it 
Department of Mathematical Sciences, University of Liverpool, Liverpool L69 7ZL, UK 
 \\[1ex] } 
${}^b$ {\it 
School of Engineering and Applied Sciences, Rotterdam University of Applied Sciences, \\ 
G.J.\ de Jonghweg 4 - 6, 3015 GG Rotterdam, the Netherlands
 \\[1ex]
Research Centre Innovations in Care, Rotterdam University of Applied Sciences, \\ 
Postbus 25035, 3001 HA Rotterdam, the Netherlands
} 
\\[1ex] 
\bigskip
\end{center}

\subsection*{\centering Abstract}

This paper aims to shed light on what becomes of discrete torsion within heterotic orbifolds when they are resolved to smooth geometries.
Gauged Linear Sigma Models (GLSMs) possessing (0,2) worldsheet supersymmetry are employed as interpolations between them. 
This question is addressed for resolutions of the non--compact $\Cplx^3/\Intr_2\times\Intr_2$ and the compact $T^6/\Intr_2\times\Intr_2$ orbifolds to keep track of local and global aspects.  
The GLSMs associated with the non--compact orbifold with or without torsion are to a large extent equivalent: only when expressed in the same superfield basis, a field redefinition anomaly arises among them, which in the orbifold limit reproduces the discrete torsion phases.
Previously unknown, novel resolution GLSMs for $T^6/\Intr_2\times\Intr_2$ are constructed.  
The GLSM associated with the torsional compact orbifold suffers from mixed gauge anomalies, which need to be cancelled by appropriate logarithmic superfield dependent FI--terms on the worldsheet, signalling $H$--flux due to NS5--branes supported at the exceptional cycles.

\newpage 
\thispagestyle{empty} 
\tableofcontents
\newpage 
\setcounter{page}{1}

\setcounter{equation}{0}
\section{Introduction}
\label{sc:Introduction}

Given the current astrophysical, collider and cosmological data, the standard cosmological and particle physics models may provide viable parameterisation of all observational data up the Planck scale. Obtaining further insight into the basic origin of these parameters necessitates the synthesis of the gauge quantum field theories with gravity. The most developed contemporary mathematical framework to explore the gauge--gravity unification is string theory.

The consistency conditions of string theory require the existence of a finite number of degrees of freedom beyond those observed in contemporary experiments. These degrees of freedom may appear in different guises. They may be interpreted as extra target space dimensions with vector bundles, or as two dimensional fields propagating on the string worldsheet. Ultimately, the different representations may describe the same physical objects and it is vital to extract the physical characteristics, irrespective of the particular language used.

The $\Intr_2\times \Intr_2$ orbifolds of six dimensional toroidal compactifications are among the most studied string constructions to date. They have been used to derive phenomenological string models and to study how the parameters of the Standard Model may be derived from string theory, using their free fermionic~\cite{Faraggi:1989ka,Faraggi:1991jr,Faraggi:1992fa, Cleaver1999,Faraggi:2006qa,Faraggi:2017cnh} and orbifold~\cite{Lebedev:2006kn,Lebedev:2007hv,Lebedev:2008un,Blaszczyk:2009in}
realisations, and their smooth resolutions~\cite{Blaszczyk:2010db}. 
Other phenomenological interesting smooth compactifications have been investigated in {\em e.g.}~\cite{Donagi:1999ez,Donagi:2000zs,Braun2005,Braun:2005nv,Bouchard:2005ag,Anderson:2009mh,Anderson:2011ns}.
These phenomenological studies encompass supersymmetric and
non--supersymmetric string vacua~\cite{Blaszczyk:2014qoa,Abel:2015oxa,Nibbelink:2015ixa,GrootNibbelink:2015lme,Faraggi:2020wld}
with symmetric and asymmetric
boundary conditions~\cite{GrootNibbelink:2017usl,GrootNibbelink:2020dib} and the $\Intr_2\times \Intr_2$ orbifolding can enable the fixing of all of the
untwisted geometrical moduli~\cite{Faraggi:2005ib}. 

The relation between worldsheet string models and their effective field theory geometrical limits presently occupies much of the discourse in string phenomenology in the form of the so--called ``swampland program''. This program aims to address the question when does an effective field theory model of quantum gravity have an ultra--violet complete embedding in string theory, and hence can be viewed as a bottom--up approach to the study of this relation.
An alternative top--down approach seeks to find the imprint of worldsheet symmetries in the effective field theory target space models. Notable 
examples of this approach include mirror symmetry~\cite{Vafa:1994} and spinor--vector
duality~\cite{Faraggi2007a,Faraggi2007b,CatelinJullien:2008pc,Angelantonj:2010zj,Faraggi:2011aw,Faraggi:2021yck,Faraggi:2021fdr}.

The worldsheet constructions of string vacua consist of a perturbative expansion in string amplitudes. They are constrained to preserve the classical symmetries of reparameterisation and Weyl invariance, {\it i.e.}\ they are invariant under modular transformations of the worldsheet parameter, and are encoded in the one--loop partition function. The requirement of modular invariance entails that the partition function is a sum over different sectors that combine to form a modular invariant object. While most of the signs in this sum are dictated by modular invariance, some other may be arbitrary and play a vital role in determining the physical properties of the string models. In particular, the origins of mirror symmetry and spinor--vector dualities may be traced back to (generalised) discrete torsions. Discrete torsions typically arise in the worldsheet constructions as a result of multiple modding out operations. For example, we may mod out by several twists of the internal dimensions; or by identifications by translations of points in the internal compactified space; or we may combine actions of these shifts and twists. Additionally, in the heterotic--string these may be combined with an action on the gauge bundles, which results in a reduction of the gauge symmetry. The spinor--vector duality, for example, arises due to the action of Wilson lines on the gauge bundles. 

The interpretation of (generalised) discrete torsions from the geometrical effective field theory point of view is obscured as one does not have an exact partition function description in which these discrete torsion phases are present. It is therefore of interest to elucidate the manifestation of the discrete torsions in the effective field theory limit. 
If there is a discrete action on the target space, this can be accompanied with discrete torsion in the form of some non--trivial action on the $B$--field~\cite{Sharpe:2000ki,Sharpe:2000wu,Sharpe:2000tw}. 
However, in this paper we however wondered what happens to the discrete torsion between orbifold twists, if one fully resolves the orbifold so that no discrete symmetries are left on the smooth target space. 
We aim to investigate this manifestation using the Gauged Linear Sigma Model (GLSM) representation of string vacua. GLSMs provide a particularly appealing framework to explore this question, as they provide a single framework in which one can interpolate between different regimes, like the singular orbifold limit and smooth compactifications.

\subsection{Main paper objectives}

One of the central objectives of this paper is to systematically study the discrete torsion phases in smooth string compactifications using the GLSM language to bridge the gap between the orbifold CFT formulations and the effective field theory descriptions for smooth target spaces. Concretely, this program is considered for $\Intr_2\times \Intr_2$ orbifolds of free CFTs where the discrete torsion is known as the Vafa--Witten phase. 

First resolutions of the non--compact $\Cplx^3/\Intr_2\times\Intr_2$ orbifold are considered in the GLSM language. To have a particular simple context the focus is on line bundle resolutions generated by physical blowup modes, twisted string states without oscillator excitations. The precise identification of such resolution GLSMs from this data was worked out in the past~\cite{GrootNibbelink:2010qut}. Since only the standard embedding bundles would allow for a $(2,2)$ worldsheet description, the incorporation of line bundles requires a $(0,2)$ GLSM language. For both orbifold CFTs without and with torsion the corresponding resolution GLSMs are constructed. In order to compare them at the Lagrangian level on the worldsheet, one has to ensure that one uses the same superfield basis. (In the path integral formulation it only make sense to compare theories using their classical actions when the same integration field variables are employed.) Hence, as the charges of the superfields in the GLSMs of the non--torsion and the torsion orbifolds do not agree, superfield redefinitions are needed before this comparison is possible. As a cross check of the applied methods the GLSMs are considered in the deep orbifold regime to investigate how the torsion phases may be recovered. 

The study of compact models with torsion is particularly intriguing since certain fluxes cannot be pushed to infinity and thereby out of the realm of the used description. Hence, the second part of the paper focusses on resolutions of compact $T^6/\Intr_2\times\Intr_2$ orbifolds without or with discrete torsion switched on. Before, a careful study of the imprints of discrete torsion can be investigated, first GLSMs for resolutions of $T^6/\Intr_2\times\Intr_2$ have to be set up. In the past GLSMs for compact orbifold resolutions were worked out in~\cite{Blaszczyk:2011hs}. Even though the necessary techniques were developed there, GLSM resolutions of $T^6/\Intr_2\times\Intr_2$ were not considered explicitly. Moreover, that paper used the $(2,2)$ language throughout. However, to match up with the considerations of the non--compact cases, it is necessary to describe resolutions of $T^6/\Intr_2\times\Intr_2$ here using $(0,2)$ GLSM terminology. Having fixed the geometrical aspects in the GLSM description, similar blowups are considered induced by non--oscillator twisted states as in the non--compact context. However, for the compact GLSM resolutions this leads to more complicated bundle constructions which take features of standard embedding bundles on the underlying torus cycles mixed with line bundles on the resolved $\Intr_2$--singularities. With all this in place, the resolution GLSMs of the compact orbifolds without and with torsion can be investigated.

\subsection{Paper organisation}

The main part of the paper starts with a short review in Section~\ref{sc:Z22orbifolds} of some features of $\Intr_2\times\Intr_2$ orbifolds to provide the necessary foundation for the subsequent investigations. Section~\ref{sc:GLSMbasics} summarises some essential prerequisites about $(0,2)$ GLSMs without which the remainder of this manuscript might be a bit hard to follow for non--experts. Further technical details on this topic are diverted to Appendix~\ref{sc:(0,2)models}. Next, Section~\ref{sc:C3Z22glsms} focusses on GLSM resolutions of non--compact $\Cplx^3/\Intr_2\times\Intr_2$ without and with torsion. 
Some properties described there rely on charge matrices which are collected in Appendix~\ref{sc:ChargeMatrices} not to interrupt the main flow of this section. Section~\ref{sc:T6Z22glsms} repeats these exercises for compact $T^6/\Intr_2\times\Intr_2$ GLSM resolutions focussing on the additional features and complications that compactness brings. Appendix~\ref{sc:Anomalies} derives gauge anomalies in two dimensions and provide  $(0,2)$ superspace expressions for them which are used frequently in Sections~\ref{sc:C3Z22glsms} and~\ref{sc:T6Z22glsms}.

\setcounter{equation}{0}
\section{Properties of $\boldsymbol{\Intr_2\times\Intr_2}$ orbifolds}
\label{sc:Z22orbifolds}

The purpose of the present section is to recall some crucial information about heterotic $\Intr_2\times\Intr_2$ orbifolds to understand their resolutions using GLSM methods that are laid out in subsequent sections. Hence, it does not aim to give a complete review of heterotic orbifolds (for more comprehensive discussions see {\em e.g.}\ \cite{Dixon:1985jw,Dixon:1986jc,ibanez_88,Ibanez:1987sn,Nilles:2011aj}). In particular, properties of $\Intr_2\times\Intr_2$ orbifolds may be found in {\em e.g.}\ \cite{Forste:2004ie,Faraggi:2004rq,Donagi2004,Donagi2008,Blaszczyk:2009in}. A crucial feature of $\Intr_2\times\Intr_2$ is that they may posses discrete torsion~\cite{Vafa:1986wx,Vafa:1994}. As is recalled here this feature determines which twisted states survive the orbifold projections.

\subsection{Orbifold twists and gauge shift vectors}

The bosonic description of the $\Intr_2\times\Intr_2$ orbifold starts with the introduction of  two twist vectors 
\equ{
v_{1}=\big(0,0, \sfrac{1}{2},\sm\sfrac{1}{2}\big)~,
\qquad 
v_{2}=\big(0,\sm\sfrac{1}{2},0, \sfrac{1}{2}\big)~, 
}
which act on the complex coordinate fields $z_u$ with $u=0,1,2,3$. Here $z_0$ denotes the four dimensional non--compact directions in light--cone gauge. (Since the main interest is on the internal coordinates, $u$ is taken to label the internal coordinates and then runs over $u=1,2,3$ only.) Thus the first entries of the twist vectors indicate that the twists act trivially on the four dimensional Minkowski space. For the non--compact orbifold $\Cplx^3/\Intr_2\times\Intr_2$ the coordinates $z_u\in\Cplx$ parametrise three complex planes. While for the compact orbifold $T^6/\Intr_2\times\Intr_2$ they parametrise the three underlying two--tori $T^2$. An arbitrary element $g$ of the $\Intr_2\times\Intr_2$ orbifold point group then corresponds to the twist vector 
\equ{
v_{g}= t_{1}\, v_{1}+t_{2}\, v_{2}~,
}
where $t_1, t_2 = 0,1$ label its four elements.

To complete the definition of the orbifold actions gauge shift vectors have to be given. In the orbifold standard embedding the gauge shift vectors are taken to be equal to these twist vectors augmented with the appropriate number of zero entries: 
\equ{ \label{eq:SpecificGaugeShifts} 
V_{1}=\big(0, \sfrac{1}{2},\sm\sfrac{1}{2}, 0^{5}\big)\big(0^{8}\big)~,
\qquad 
V_{2}=\big(\sm\sfrac{1}{2},0, \sfrac{1}{2}, 0^{5}\big)\big(0^{8}\big)~, 
}
and define the gauge shift embedding
\equ{
V_{g}=t_{1}\, V_{1}+t_{2}\, V_{2}~, 
}
for each of the four orbifold point group elements. As the notation of the shift vectors suggest, this paper uses the $E_8\times E_8$ heterotic string for concreteness. In addition a heterotic orbifold might feature a number of discrete Wilson lines. In this paper the consequences of them are not considered.

\subsection{Discrete torsion phase} 
\label{sc:DiscreteTorsionPhase}

At the one loop level it is conventional to distinguish between constructing elements $g, h$ of the orbifold group, which define the different orbifold sectors of the theory, and the projecting elements $g', h'$, which implement the appropriate orbifold projections. Hence, on the one loop worldsheet torus a heterotic orbifold model is defined uniquely by the properties introduced above up to a possible discrete torsion phase~\cite{Vafa:1986wx,Vafa:1994}
\equ{
\gF^{\!\times}{}^{t_1,t_2}_{t_1',t_2'} = e^{\pi i\, \gve^{\!\times} (t_{1}^{\phantom{\prime}} t_{2}^{\prime}-t_{2}^{\phantom{\prime}} t_{1}^{\prime})}~
}
in its one loop partition function~\cite{Vafa:1986wx}. The possible torsion phase leads to a specific interplay between the constructing and projecting orbifold group elements. Clearly, if $\gve^{\!\times}=0$ there is no torsion as the torsion phase is equal to unity, but if $\gve^{\!\times}=1$ the model possesses discrete torsion as the phase is non--trivial. 

An alternative equivalent way that discrete torsion can be introduced is by so--called brother models, {\em i.e.}\ models with gauge shift vectors that differ from the original ones by appropriate lattice vectors~\cite{Ploger:2007iq}. In particular, for the model~\eqref{eq:SpecificGaugeShifts} the brother model has gauge shift vectors 
\equ{ \label{eq:BrotherGaugeShifts} 
V^{\!\times}_{1}= -V_1=\big(0, \sm\sfrac{1}{2},\sfrac{1}{2}, 0^{5}\big)\big(0^{8}\big)~,
\qquad 
V^{\!\times}_{2}=-V_2=\big(\sfrac{1}{2},0, \sm\sfrac{1}{2}, 0^{5}\big)\big(0^{8}\big)~, 
}
so that their differences are indeed lattice vectors.

\subsection{Orbifold spectra with(out) torsion}
\label{sc:OrbifoldSpectraWithoutTorsion}

Any state in the orbifold spectrum may be characterised by two shifted momenta 
\equ{ \label{eq:ShiftedMomenta} 
p_{g}=p+v_{g}~, 
\qquad 
P_{g}=P+V_{g}~, 
}
where the vector $p$ is an element of the lattice $V_4\oplus S_4$ and $P$ of $(O_8\oplus S_8)\otimes (O_8\oplus S_8)$. The shifted momenta of level matched massless states are subject to the following two conditions 
\equ{ 
\frac{1}{2}\,p_{g}^{2} = \frac{1}{2}-\delta c_g~, 
\qquad 
\frac{1}{2}\, P_{g}^{2} = 1-\delta c_g- \go_{g} \cdot \widetilde{N}_{g}-\bgo_{g} \cdot \overline{\widetilde{N}}_{g}~,  
}
where the orbifold vacuum shift 
\equ{
\delta c_g = \frac{1}{2} \sum_u \go_{g,u}(1-\go_{g,u})
}
is defined in terms of $\go_{g,u} \equiv (v_g)_u$ and $\bgo_{g,u} \equiv - (v_g)_u$ which satisfy the inequalities: $0 < \go_{g,u}, \bgo_{g,u} \leq 1$. Finally, $(\widetilde{N}_g)_u$ and $(\overline{\widetilde{N}}_g)_u$ are the number operators that count the number of right--moving oscillators act on the state. Only the states that survive the orbifold projection conditions, 
\equ{
P_{g} \cdot V_{g'}-p_{g} \cdot v_{g'} \equiv  
\frac{1}{2}\big(V_{g} \cdot V_{g'}-v_{g} \cdot v_{g'}\big) 
+ \big(\overline{\widetilde{N}}_{g}-\widetilde{N}_{g}\big) \cdot v_{g'} 
+\frac{\gve^{\!\times}}{2} \big(t_{1}^{\phantom{\prime}} t_{2}^{\prime}-t_{2}^{\phantom{\prime}} t_{1}^{\prime}\big)~, 
}
are part of the physical orbifold spectrum. The last term in these projection conditions encodes the consequences of discrete torsion on the massless spectrum. Consequently, the discrete torsion phases only affect the twisted sectors. The resulting orbifold spectrum is conventionally divided in a number of sectors:

\subsubsection*{Untwisted sector}

The untwisted sector is identified by $(t_1,t_2) = (0,0)$. This sector corresponds to so--called bulk states which live everywhere within the internal geometry. It contains the metric, the anti--symmetric tensor and the dilaton degrees of freedom as well as the target space gauge fields and all their superpartners in ten dimensions. The non--Abelian unbroken gauge group in four dimensions is $E_6 \times E_8$. 
In addition, there are three copies of charged matter in the $(27)+(\overline{27})$ of $E_6$ independently of whether torsion is switched on or not.

\subsubsection*{Twisted sectors} 

There are three twisted sectors with $t=(t_1,t_2): 1=(1,0), 2=(0,1)$ and $3=(1,1)$ which only posses $\cN=1$ supersymmetry in six dimensions\footnote{Also sometimes referred to as $\cN=2$ sectors from the four dimensional point of view.}: On the non--compact orbifold $\Cplx^3/\Intr_2\times\Intr_2$ the corresponding twisted states are localised at the three complex codimension two singularities of the three non--trivial orbifold twists. Each twisted sector is supported on 16 fixed two--tori within the compact orbifold $T^6/\Intr_2\times\Intr_2$. Half of these states are projected out by the orbifold action of the second orbifold element. Which half depends on whether torsion is switched on, see Table~\ref{tb:TwistedSpectrumWithOutOscillators} which gives the twisted states without twisted oscillator excitations.

\begin{table} 
\[
\begin{array}{|c||c|c|c|c|}
\hline 
\text{Sector} & \text{Shifted momentum } P_g & \text{Repr.} & \gve^{\!\times} =0 & \gve^{\!\times} = 1 
\\ \hline\hline 
 & 
\big(1,\sm\sfrac{1}{2},\sm\sfrac{1}{2}, 0^{5}\big)\big(0^{8}\big) 
& (1) & \multirow{ 2}{*}{\text{in}} & \multirow{ 2}{*}{\text{out}}
\\ \cline{2-3} 
1=& \big(\sm1,\sm\sfrac{1}{2},\sm\sfrac{1}{2} 0^{5}\big)\big(0^{8}\big); 
\big(0, \sfrac{1}{2},\sfrac{1}{2}, \underline{\pm 1,0^{4}}\big)\big(0^{8}\big); 
\big(\sm\sfrac{1}{2},0,0,\underline{\sm\sfrac{1}{2}^e,\sfrac{1}{2}^{5-e}}\big)\big(0^{8}\big) 
& (27) & & 
\\ \cline{2-5} 
(1,0) & \big(\sm1, \sfrac{1}{2}, \sfrac{1}{2}, 0^{5}\big)\big(0^{8}\big) 
& (\overline{1}) & \multirow{ 2}{*}{\text{out}} & \multirow{ 2}{*}{\text{in}} 
\\ \cline{2-3}  
& \big(1,\sfrac{1}{2},\sfrac{1}{2},0^{5}\big)\big(0^{8}\big); 
\big(0,\sm\sfrac{1}{2},\sm\sfrac{1}{2},\underline{\pm1,0^{4}}\big)\big(0^{8}\big); 
\big(\sfrac{1}{2},0,0,\underline{ \sm\sfrac{1}{2}^o,\sfrac{1}{2}^{5-o}}\big)\big(0^{8}\big)
& (\overline{27})  &  &   
\\ \hline\hline 
 & \big(\sm\sfrac{1}{2},1,\sm\sfrac{1}{2},0^{5}\big)\big(0^{8}\big)
& (1) & \multirow{ 2}{*}{\text{in}} & \multirow{ 2}{*}{\text{out}} 
\\ \cline{2-3}
2= &\big(\sm\sfrac{1}{2},\sm1,\sm\sfrac{1}{2},0^{5}\big)\big(0^{8}\big); 
\big(\sfrac{1}{2},0,\sfrac{1}{2},\underline{\pm1,0^{4}}\big)\big(0^{8}\big); 
\big(0,\sm\sfrac{1}{2},0, \underline{\sm\sfrac{1}{2}^{e},\sfrac{1}{2}^{5-e}}\big)\big(0^{8}\big)
& (27) &  & 
\\ \cline{2-5} 
(0,1) & \big(\sfrac{1}{2},\sm1,\sfrac{1}{2},0^{5}\big)\big(0^{8}\big)
& (\overline{1}) & \multirow{ 2}{*}{\text{out}} & \multirow{ 2}{*}{\text{in}} 
\\ \cline{2-3} 
& \big(\sfrac{1}{2},1,\sfrac{1}{2},0^{5}\big)\big(0^{8}\big);
\big(\sm\sfrac{1}{2},0,\sm\sfrac{1}{2},\underline{\pm1,0^{4}}\big)\big(0^{8}\big);
\big(0,\sfrac{1}{2},0,\underline{\sm\sfrac{1}{2}^o,\sfrac{1}{2}^{5-o}}\big)\big(0^{8}\big)
& (\overline{27}) &  & 
\\ \hline\hline 
 &\big(\sm\sfrac{1}{2},\sm\sfrac{1}{2},1,0^{5}\big)\big(0^{0}\big) 
& (1) & \multirow{ 2}{*}{\text{in}} & \multirow{ 2}{*}{\text{out}}
\\ \cline{2-3}
3= &\big(\sm\sfrac{1}{2},\sm\sfrac{1}{2},\sm1,0^{5}\big)\big(0^{8}\big);
\big(\sfrac{1}{2},\sfrac{1}{2},0,\underline{\pm1,0^{4}}\big)\big(0^{8}\big); 
\big(0,0,\sm\sfrac{1}{2},\underline{\sm\sfrac{1}{2}^{e},\sfrac{1}{2}^{5-e}}\big)\big(0^8\big) 
& (27) &  & 
\\ \cline{2-5} 
(1,1) &\big(\sfrac{1}{2}, \sfrac{1}{2},\sm1,0^{5}\big)\big(0^8\big) 
& (\overline{1})  & \multirow{ 2}{*}{\text{out}} & \multirow{ 2}{*}{\text{in}}
\\ \cline{2-3}
&\big(\sfrac{1}{2},\sfrac{1}{2},1,0^{5}\big)\big(0^{8}\big); 
\big(\sm\sfrac{1}{2},\sm\sfrac{1}{2},0,\underline{\pm1,0^{4}}\big)\big(0^{8}\big); 
\big(0,0,\sfrac{1}{2},\underline{\sm\sfrac{1}{2}^o,\sfrac{1}{2}^{5-o}}\big)\big(0^8\big) 
& (\overline{27})  & &  
\\ \hline
\end{array} 
\]
\caption{\label{tb:TwistedSpectrumWithOutOscillators}
This table lists the twisted sector spectra obtained from non--oscillator excitation states and indicates whether they are in the physical spectrum without or with torsion, $\ge^{\!\times} =0$ or $1$, respectively. }
\end{table}

\setcounter{equation}{0}
\section{Geometries and bundles from $\boldsymbol{(0,2)}$ gauged (linear) sigma models}
\label{sc:GLSMbasics}

\begin{table}
\[
\arry{|c||c|c|c
||c|c||c|c||c|c|c||c|c|}{
\hline
\text{Superfield} & \der & \bder & D_+
& \gF^a & \gL^m & \gPs^A & \gG^M & V_i & A_i & F_i & \gS_I & \gY_I 
\\ \hline 
\text{Phys.\,Comp.} &&&& (z^a,\gf^a) &  (\gl^m,h^m) & (y^A,\gps^A) &  (\gg^M, f^M)  & \multicolumn{3}{c||}{(A_\gs^i, A_\bgs^i, \gvf^i, D^i)} & \multicolumn{2}{c|}{(s^I, \gch^I)} 
\\ \hline 
\#  &&&& \mathrm{N}_\gF & \mathrm{N}_\gL & \mathrm{N}_\gPs & \mathrm{N}_\gG & \multicolumn{3}{c||}{\mathrm{N}_V} & \multicolumn{2}{c|}{\mathrm{N}_\gS} 
\\ \hline\hline 
\cL & 0 & 1 & 0  
& 0 & \sfrac 12 & 0 & \sfrac 12 & 0 & 1 & 1 & \sfrac 12 & \sfrac 12 
\\ \hline 
\cR & 1 & 0 & \sfrac 12 
& 0 & 0 & 0 & 0 & 0 & 0 & \sfrac 12 & 0 & \sfrac 12 
\\ \hline 
\text{R} & 0 & 0 & -1
&  0 & 0 & 1 & 1 & 0 & 0 & 1 & 0 & 1 
\\ \hline\hline 
\cQ & 0 & 0 & 0 
& (q_i)^a & (Q_i)^m & (\textsf{q}_i)^A & (\textsf{Q}_i)^M & \text{n.l.} & \text{n.l.} & 0 & 0 & 0
\\ \hline 
}
\]
\caption{\label{tb:Superfields} 
This table specifies the left-- and right--Weyl dimensions, $\cL$ and $\cR$, the $\text{R}$--charge and the gauge charges $\cQ_i$ of the operators $\der, \bder, D_\pm$ and the superfields which may be used in a (0,2) GLSM. The physical components of these multiplets are indicated as well as the indices that label them; the third line gives the total number of these multiplets. 
}
\end{table}

\subsection{(0,2) Superfields} 
\label{sc:Superfields}

Two dimensional theories with $(0,2)$ supersymmetry admit a number of different types of superfields (or multiplets). Appendix~\ref{sc:(0,2)models} gives a short review of $(0,2)$ superfields on superspace and sets notations and conventions used in this work. Gauged sigma models are a special class of $(0,2)$ theories with bosonic and possibly also fermionic gaugings. The superfields used in this work are summarised in Table~\ref{tb:Superfields} and the labels used to enumerate them are indicated there. In addition, their gauge charges, left-- and right--Weyl dimensions and R--charges (defined in Appendix~\ref{sc:SCtransformations}) are given.

The most important matter superfields are chiral and chiral Fermi multiplets. A chiral multiplet $\gF = (z, \gf)$ contain a complex scalar $z$ and a right--moving fermion $\gf$. A chiral Fermi multiplet $\gL = (\gl, h)$ consists of a left--moving fermion $\gl$ and an auxiliary scalar field $h$. In addition, there are chiral multiplets $\gPs =(y, \gps)$ and chiral Fermi multlplets $\gG=(\gg,f)$. The distinction between these chiral and chiral Fermi superfields is made by their $\text{R}$--symmetry charge: $\gF$ and $\gL$ are neutral while $\gPs$ and $\gG$ carry charge 1. The last line of this table gives the gauge charges and dictates the super gauge transformations of these matter superfields.

For the corresponding bosonic gaugings vector multiplets have to be introduced consisting of two real bosonic superfields $V$ and $A$ from which gauge invariant super field strengths $F$ can be constructed 
\equ{
F = - \sfrac 12 \bD_+ \big(A - i \bder V\big)~. 
}
The physical components of these multiplets are the gauge field $A_\gs$, $A_\bgs$ with field strength $F_{\gs\bgs} = \der_\gs A_\bgs - \der_\bgs A_\gs$ and a right--moving fermion $\gvf$ and a real auxiliary field $D$.

On the chiral Fermi multiplets fermionic gauge transformations 
\equ{ \label{eq:FermionicGaugeTrans} 
\gL \ra \gL + U(\gF) \!\cdot\! \gX~, 
\qquad 
\gG \ra \gG + \gPs\,W(\gF) \!\cdot\! \gX
}
may act with chiral Fermi super gauge parameters. To obtain invariant action under these transformation, Fermi gauge multiplets $\gS$ need to be introduced with super field strengths
\equ{
\gY =  \bD_+ \gS~. 
}
Their physical components are complex scalars $s$ and left--moving fermions $\gch$.

A few comments are in order. The theories that are studied here do not define proper string theories as their worldsheet actions are not fully conformal. In particular, dynamical gauge fields on the worldsheet are not scale invariant as their gauge coupling is dimensionfull. Nevertheless it is useful to use characterisations, like the left-- and right--moving Weyl dimensions, as in the scale invariant limit the corresponding superconformal symmetries are recovered. Moreover, the ``linear'' in GLSMs signifies that only kinetic terms quadratic in the fields are considered, while in non--linear sigma models this restriction is lifted for chiral superfields.

The main reason why GLSMs are of interest for string theory is that they can provide interesting insights in how geometries and vector bundles on them can arise:

\subsection{Emergent effective geometry}

The scalar part of GLSMs can be associated to target space geometries like weighted projective spaces, complete intersection Calabi--Yaus and many generalisations of these as was realised by the pioneering paper~\cite{Witten:1993yc}. The scalar components $z$ of the chiral multiplets $\gF$ can be interpreted as the homogeneous coordinates of projected spaces, where the $\Cplx^*$--scalings are encoded by the scalar part of the super gauge transformations: 
\equ{
z \ra e^{q_i\cdot \gth}\, z~, 
\qquad  
\gth = \sfrac 12\, a - i\, \ga \in \Cplx^{N_V}~.
} 
In the Wess--Zumino gauge the sizes of these projective spaces are set by the D--term equations 
\equ{ \label{eq:Dterms} 
 \sum_a (q_i)^{a} |z^a|^2  
 = r_i~, 
}
for each $i = 1,\ldots, N_V$. (In principle there is a second sum over the scalars $y^A$ here, but they are typically all forced to zero as discussed below.) Here the parameters $r$ are the real parts of the Fayet--Iliopoulos (FI) coefficients $\gr(\gF)$ which define superpotentials involving the super gauge field strengths 
\equ{ \label{eq:FISuperpotential} 
W_\text{FI} = \gr(\gF)\!\cdot\! F~, 
\qquad 
\gr(z) = \sfrac 12\, r + i\, \gb \in \Cplx^{N_V}~.  
}
This is gauge invariant if the functions $\gr(\gF)$ are neutral. The target space interpretation of $r$ are moduli, that set the radii of certain cycles, and $\gb$ may be interpreted as axions in the effective geometry.

String backgrounds, like Calabi--Yaus, are often defined as hypersurfaces in such projected spaces. In the GLSM language this can be encoded in a $(0,2)$ superpotential 
\equ{ \label{eq:GeomSuperpotential}
P_\text{geom} =  \gG\, P(\gF)~. 
}
In the conformal limit, the scalar components of the algebraic equations of motion of chiral Fermi superfields $\gG^M$ lead to F--term equations:
\equ{ \label{eq:Fterms} 
P_M(z) = 0~,
}
for $M=1,\ldots, N_\gG$, which precisely cut out such hypersurfaces. Consequently, the dimension of the resulting target space manifold $\cM$ equals: 
\equ{
\text{dim}_\Cplx(\cM) = \mathrm{N}_\gF - \mathrm{N}_V - \mathrm{N}_\gG~. 
}
This should be equal to $2$ or $3$ if one only considers the internal manifold of complex dimension $2$ or $3$; or $4$ if the complete spacetime in light--cone gauge is described by the GLSM.

In addition, the GLSM description can be used to determine an atlas of coordinate patches: in a given phase one or multiplet set(s) of scalar fields are necessarily non--zero. Hence, by analysing the combined D--term and F--term equations, \eqref{eq:Dterms} and~\eqref{eq:Fterms}, all the coordinate patches within a phase of the GLSM can be determined.

\subsection{Emergent effective vector bundle}

The part of $(0,2)$ GLSMs that involve the chiral Fermi multiplets can be interpreted as vector bundles (or as sheafs if they are not fully regular)~\cite{Witten:1993yc,Distler:1996tj,Chiang:1997kt}. The fermionic components $\gl$ of the Fermi multiplets $\gL$ are line bundle sections on this manifold as their $\Cplx^*$--scalings read 
\equ{
\gl \ra e^{Q_i\cdot \gth}\, \gl~. 
} 
If there are no fermionic super gauge transformations and no chiral superfields $\gPs$ in the model, then the target space gauge background is simply a collection of line bundles. 

However, in general, they describe a more complicated vector bundle $\cV$ which is derived from a complex (generalisation of a monad construction), since they have to satisfy the constraints 
\equ{ \label{eq:BundleFterms}
M(z) \gl = 0~, 
}
due to the lowest components of the algebraic equations of motion of $\gPs$ that follows from the bundle superpotential 
\equ{ \label{eq:BundleSuperpotential} 
P_\text{bundle} = \gPs\, M(\gF)\, \gL
}
and are subject to gauge transformations 
\equ{ \label{eq:BundleInjections} 
\gl \ra \gl + U(z)\!\cdot\! \gx~, 
}
which are the lowest components of the fermionic super gauge transformations~\eqref{eq:FermionicGaugeTrans}. Combined the equations~\eqref{eq:BundleFterms} and~\eqref{eq:BundleInjections}  imply that a vector bundle $\cV = \text{Ker}(U)/\text{Im}(M)$ is constructed from the complex
\equ{  
0 \ra \cO^{ \mathrm{N}_\gS} \stackrel{U}{\lra} \bigoplus\limits^{\widetilde{\mathrm{N}}_\gL}_{m=1} \cO(Q^m) \stackrel{M}{\lra} \bigoplus\limits^{\mathrm{N}_\gPs}_{A=1} \cO(-\textsf{q}^A) \ra 0~.  
}
Here $\widetilde{\mathrm{N}}_\gL\leq {\mathrm{N}}_\gL$ denotes the number of interacting Fermi multiplets in the GLSM. (The numbers in the $\cO$s of such complexes are conventionally integers. But in the normalisations used in this paper they might be fractional (like $1/2$), hence they should then be multiplied by an appropriate common factor. In addition, the charges of the chiral superfields $\gPs$ are negative in the conventions used in this work and they set the degrees of the constraints~\eqref{eq:BundleFterms} on the fermions.) The dimensionality of the fibers of resulting vector bundle $\cV$ is given by 
\equ{
\text{dim}_\Cplx(\cV) = \widetilde{\mathrm{N}}_\gL - \mathrm{N}_\gS - \mathrm{N}_\gPs~, 
}
provided that  $M(z)$ and $U(z)$ have maximal ranks $\mathrm{N}_\gPs \leq \mathrm{N}_\gL$ and $\mathrm{N}_\gS \leq \mathrm{N}_\gL$, respectively~\cite{Distler:1996tj}. (If this is not everywhere the case, this indicates that there are singularities in the bundle  instead.) In order that this bundle can be embedded in the gauge degrees of freedom of the heterotic string $\text{dim}_\Cplx(\cV)$ should less than eight so as to fit within an $E_8$--factor. (The bundle might also fill up part of both $E_8$--factors, but then it has to split accordingly.) Since the full rank of $E_8\times E_8$ is 16, the total number of Fermi multiplets is given by ${\mathrm{N}}_\gL = 16+ \mathrm{N}_\gS +\mathrm{N}_\gPs$. Hence, there are a number of spectator (non--interacting and neutral) Fermi multiplets $\gL_n$, $n=1,\ldots, {\mathrm{N}}_\gL-\widetilde{\mathrm{N}}_\gL$, which lead to the unbroken gauge degrees of freedom in target space.

The superpotential~\eqref{eq:BundleSuperpotential} has another important consequence: If $M(z)$ has maximal rank, the equations of motion of $\gL$ induced by the bundle superpotential~\eqref{eq:BundleSuperpotential} imply that all $y^A = 0$. This was implicitly assumed when~\eqref{eq:Dterms} were written down, since, in general, also contributions from the scalars $y^A$ should be present in these equations.

The fermionic gauge transformations~\eqref{eq:FermionicGaugeTrans} only leaves the superpotentials~\eqref{eq:GeomSuperpotential} and~\eqref{eq:BundleSuperpotential} combined inert when the following compatibility conditions hold
\equ{ \label{eq:SPfermigaugeInv}
W_A{}^{IM}(\gF) P_M(\gF) + M_{Am}(\gF) U^{mI}(\gF) = 0~. 
}
In general, it is not so straightforward to find functions such that these conditions are fulfilled. However, when the superpotentials and the fermionic gaugings are taken to lie on the $(2,2)$ locus discussed below, these conditions are automatically satisfied.

\subsection{The (2,2) locus}
\label{sc:22Locus} 

The interacting part of $(0,2)$ GLSMs (or at least the part that involves fermionic gaugings) might possess a higher amount of supersymmetry. For this to happen the (0,2) multiplets need to be able to pair up. This means in particular, that there are the following relations between the number of interacting multiplets:
\equ{ 
\widetilde{\mathrm{N}}_\gL = \mathrm{N}_\gF~, 
\qquad
\mathrm{N}_\gG = \mathrm{N}_\gPs~, 
\qquad 
\mathrm{N}_\gS = \mathrm{N}_V~, 
}
This allows to identify various indices: $m=a$, $M=A$ and $I=i$; we use the latter indices for each type of indices. Furthermore, the gauge charges of chiral and Fermi multiplets need to line up: 
\equ{ 
Q_i = q_i~,
\qquad 
\textsf{Q}_i = \textsf{q}_i~. 
}
When some of these relations are not satisfied it is impossible to deform the interactions of the $(0,2)$ GLSM to become $(2,2)$. If this is possible, then the $(0,2)$ theory is said to be on the $(2,2)$ locus. 

On the $(2,2)$ locus of the space of $(0,2)$ GLSM, exact $(2,2)$ models possess various interactions encoded in the various functions introduced that need to be of a very specific form. The relations given here are subject to specific normalizations; but the implied proportionalities are essential. First of all, the functions $U(\gF)$ and $W(\gF)$ that describe the Fermi gauge transformations now read 
\equ{ 
U^{ai}(\gF) = (q_i)^a \gF^a~, 
\qquad 
W_{A}{}^{iB} = (\textsf{q}_i)^B \gd^B_A~. 
} 
They are fully dictated by the index structure and the gauge charges $(q_i)^a$ and $(\textsf{q}_i)^A$.  The functions $M(\gF)$ are determined as the derivatives of $P(\gF)$:
\equ{ \label{sc:StandardEmbeddingSuperpotential}
M_{Aa}(\gF) = P_{A,a}(\gF)~, 
}
where $F_{,a}(\gF)$ denotes the partial derivative of $F(\gF)$ with respect to $\gF^a$. Consequently, the invariance of the superpotential action under fermionic gauge transformations~\eqref{eq:SPfermigaugeInv} reduces to the gauge invariance of the superpotential: 
\equ{ 
(\textsf{q}_i)^A P_A(\gF) + P_{A,a}(\gF) \gF^a (q_i)^a = 0~. 
} 
%

\subsection{Worldsheet instantons and flux quantisation}

It is possible that on the worldsheet non--trivial gauge configurations, like instantons, are realised. The involved gauge fluxes need to be properly quantised~\cite{Adams:2009tt}: 
\equ{ \label{eq:FluxQuantisation} 
 \sum_{j}\left (q_{j}  \right )^{a} \int \frac{F_{E_{2}}^{j}}{2\pi}\in\Intr~, 
 \qquad 
  \sum_{j}\left (\mathsf{q}_{j}  \right )^{A} \int \frac{F_{E_{2}}^{j}}{2\pi}\in\Intr 
}
for all charged chiral superfields $\gF^a$ and $\gPs^A$. Here the subscript $E$ indicates that the gauge fluxes are computed in the Euclidean theory.

\subsection{Anomaly consistency conditions}
\label{sc:AnomalyConsistencyConditions}

On a GLSM there are a number of requirements in order that the theory is both consistent as a quantum theory and that it is likely to have the right properties in the conformal limit.

First of all, like any gauge theory, the GLSM has to be free of gauge anomalies. With the gauge charges given in Table~\ref{tb:Superfields}, this amounts to the following conditions 
\equ{ \label{eq:GaugeAnomalies} 
\cA_{ij} = 
-\sum_a (q_i)^a (q_j)^a  - \sum_A (\textsf{q}_i)^A(\textsf{q}_j)^A + \sum_m  (Q_i)^m (Q_j)^m + \sum_M  (\textsf{Q}_i)^M  (\textsf{Q}_j)^M  \stackrel{!}{=} 0~, 
}
for all $i,j = 1,\ldots, N_V$. The signs in these equations are determined by whether the fermions in the matter multiplets are right-- or left--moving. For $j=i$ this corresponds to pure and for $j\neq i$ to mixed gauge anomalies.

The left--, right--Weyl dimensions and $\text{R}$--charge correspond to bosonic parts of super conformal symmetries in the scale invariant limit of the GLSM. For this limit not to be obstructed the mixed left-- and right--Weyl gauge anomalies should vanish.  In detail, from Table~\ref{tb:Superfields} it follows that the left--Weyl -- gauge anomalies vanish provided that 
\equ{ \label{eq:leftWeylAnomalies} 
\sum_m (Q_i)^m  + \sum_M (\textsf{Q}_i)^M  \stackrel{!}{=} 0~, 
}
for all $i$, since the only charged superfields that carry $\cL$--charge are $\gL$ and $\gG$. These conditions can be summarised by the demand that the sum of the charges of all chiral Fermi superfields need to vanish for each gauge symmetry separately.

In addition, the charged right--moving fermions $\gf$ and $\gg$ are obtained by hitting chiral multiplets $\gF$ and $\gPs$ with $D_+$, hence the right--Weyl -- gauge anomalies are absent when 
\equ{ \label{eq:rightWeylAnomalies} 
\sum_A (\textsf{q}_i)^a + \sum_A (\textsf{q}_i)^A  \stackrel{!}{=} 0~, 
}
for all $i$. Thus, these conditions say that the sum of the charges of all chiral superfields need to vanish for each gauge symmetry separately. At the same time these conditions ensure that the FI--parameters~\eqref{eq:FISuperpotential} do not renormalise. If this isn't the case, it would not be possible to interpret them to set the scales of target space cycles as they would always run off to zero or infinity.

Finally, the $\text{R}$--symmetry survives quantisation provided that 
\equ{ \label{eq:RAnomalies} 
\sum_a (q_i)^a  + \sum_M  (\textsf{Q}_i)^M  \stackrel{!}{=} 0~, 
}
for all $i$, since the right--moving fermions $\gf$  and the left--moving fermions $\gg$ have $\text{R}$--charges $-1$ and $+1$, respectively, and opposite chiralities. When these equations are combined with~\eqref{eq:rightWeylAnomalies}, they can be stated as the sum of the charges of the chiral Fermi superfields $\gG$ have to be equal to that of the chiral superfields $\gPs$.

%
%
%

\subsection{Worldsheet Green--Schwarz mechanism: Torsion and NS5--branes}

When the gauge anomalies do not vanish, {\em i.e.}\ not all $\cA_{ij}$ in~\eqref{eq:GaugeAnomalies} vanish, the GLSM is anomalous. It is sometimes possible that certain field dependent none gauge invariant FI--terms~\eqref{eq:FISuperpotential} are precisely able to cancel these gauge anomalies~\cite{Blaszczyk:2011ib,Quigley:2011pv}. The FI--term coefficients $\gr(\gF)$ then need to transform as a shift under the anomalous gauge symmetries. This can be viewed as a Green--Schwarz mechanism on the worldsheet and might have some far reaching consequences for the geometry and the interpretation of the theory.

To understand how this comes about, note that in the naive conformal limit, the kinetic terms of the vector multiplets $V, A$ can be set to zero and their equations of motion become non--dynamical. In particular, the superfields $A$ appear linear in the actions of the chiral multiplets~\eqref{eq:GLSM_chiral} and the FI--terms~\eqref{eq:GLSM_FI}, hence their equation of motion lead to superfield constraints: 
\equ{ \label{eq:ConstraintV} 
\bgF \,e^{2q\cdot V} \!q_i\, \gF =  \gr_i(\gF) + \bgr_i(\bgF)~. 
}
Thus after enforcing the equations of motion of $A$, the vector multiplets $V$ become (implicit) functions of the chiral superfields $\gF$ and their conjugates $\bgF$. In the Wess--Zumino gauge the lowest component of these equations are the $D$--term constraints~\eqref{eq:Dterms}. However, in any gauge from~\eqref{eq:ConstraintV} it can be inferred which (scalars of the) chiral multiplets are necessarily non--zero in a given phase with a certain choice of the FI--parameters. Hence, a unitary gauge can be chosen such that all chiral superfields, that are necessarily non--zero, are set to such values that the solution for the vector superfields $V$ are all zero when all of the remaining chiral superfields are vanishing\footnote{In the remainder of this paper for presentational simplicity, the D--term equations~\eqref{eq:Dterms} are given in the Wess--Zumino gauge, while for the analysis of the torsional effects~\eqref{eq:ConstraintV} the unitary gauges, as defined here, are used implicitly.}.

Non--constant FI--terms~\eqref{eq:FISuperpotential} modify the target space geometry and generically introduces torsion onto it in the form of non--vanishing $H$--flux~\cite{Adams:2006kb,Adams:2009tt,Adams:2012sh}. Indeed, since by~\eqref{eq:ConstraintV} the vector superfields $V$ become (implicit) functions of the chiral multiplets. Inserting them in the kinetic terms of the chiral multiplets shows that the torsion tensor, the three--form $H$, 
\equ{ \label{eq:Hflux} 
H_{ab\uc} \sim \gr_{,[a}\!\cdot\! V_{,b]\uc}~, 
}
is non--zero in general, see Appendix~\ref{sc:FromGtoNLSM} or ref.~\cite{Quigley:2011pv} for a derivation. (It reads here {\em in general}, because if both $\gr_i$ and $V_i$ only depend on a single chiral superfield this expression still anti--symmetrises to zero.)
Since typically, the GLSM only contains chiral superfields $\gF$, that are linearly charged under the gauge symmetries, the required FI--coefficients can only be made by taking logarithms of combinations of them. As was argued in~\cite{Blaszczyk:2011ib,Quigley:2011pv,Melnikov:2012nm} such logarithmic singularities can be viewed as the imprints of non--perturbative physics in the form of NS5--branes on the worldsheet as the target space exterior derivative of~\eqref{eq:Hflux} lead to delta--function--like sources in the Bianchi identity of the three--form\footnote{In addition, the inclusion of log--dependent FI--terms may lead to a back reaction to the geometry~\cite{Quigley:2011pv,Melnikov:2012nm}; in this paper these consequences are not studied in detail.}.

\subsection{Orbifold resolution GLSMs}

Even though this section so far described properties of GLSMs in general, the main focus of this work is on GLSMs which are associated to (toroidal) orbifold resolutions. 
The study of resolution of singularities using $(0,2)$ GLSMs have a long history. Some pioneering works are~\cite{Distler:1996tj,Chiang:1997kt}. 
A GLSM orbifold resolution construction has the advantage over other methods to match the singular orbifold situations for which exact CFT descriptions exists with smooth compactifications using effective field theory methods. Within a single GLSM framework one has both access to the orbifold phase as well as completely resolved (and potentially many other) phases. The trade off here is that a GLSM is not (yet) a full blown CFT description. 

A fully complete correspondence between orbifold CFTs and GLSMs does not exists, but two methods have been uncovered in the past which apply to partially overlapping situations: 
\enums{
\item[{\bf A}] {\bf Twisted shifted momenta as (0,2) GLSM charges~\cite{GrootNibbelink:2010qut}:}
\\[1ex]
As was recalled in Section~\ref{sc:OrbifoldSpectraWithoutTorsion}, twisted states are uniquely identified by their shifted right-- and left--moving momenta~\eqref{eq:ShiftedMomenta}. In particular,  the right-- and left-moving shifted momenta of non--oscillator massless twisted states automatically satisfy the pure anomaly cancellation conditions when they are interpreted as GLSM gauge charges of chiral and chiral Fermi superfields, respectively. In target space these configurations may have the interpretation of line bundles on the resolved local singularities. 
\item[{\bf B}] {\bf (2,2) GLSMs for toroidal orbifold resolutions~\cite{Blaszczyk:2011hs}:}
\\[1ex]
Contrary, full global orbifold resolutions in the standard embedding can be obtained in $(2,2)$ GLSMs. The underlying two--tori are described using (variants of) the Weierstrass models. On some of their homogeneous coordinates additional (exceptional) gaugings are implemented.  For certain ranges of their FI--parameters the fixed point structure of toroidal orbifolds, while for others resolved compact Calabi--Yaus emerge. 
}
In the next section method A is employed, while in Section~\ref{sc:T6Z22glsms}
method A is combined with a partial (0,2) reduction of method B for the case of $T^6/\Intr_2\times\Intr_2$ orbifold resolutions that were not discussed in the literature before explicitly.

\setcounter{equation}{0}
\section{Non--compact $\boldsymbol{\Cplx^3/\Intr_2\times\Intr_2}$ resolution GLSMs}
\label{sc:C3Z22glsms}

\begin{table} 
\[
\arry{|c||c|c|c||c|c|c||c||c|c|c|}{
\hline 
\text{Superfield} & \gF_{1} & \gF_{2} & \gF_{3} & \gF_{1}' & \gF_{2}' & \gF_{3}' & \gL=(\gL^1,\dots,\gL^{16}) & {\gO_1} & {\gO_2} & {\gO_3} 
\\ \hline 
\text{U(1) charge} & z_{1} & z_{2} & z_{3}  & x_{1} & x_{2} & x_{3} & \gl=(\gl^1,\ldots,\gl^{16}) & {\go_1} & {\go_2} & {\go_3} 
\\ \hline\hline 
E_{1} & 0 & \sfrac 12 & \sfrac 12 & -1 & 0 & 0 & Q_1 = (Q_1^1,\ldots,Q_1^{16}) & 1 & 0 & 0  
\\ \hline 
E_{2} & \sfrac 12 & 0 & \sfrac 12 & 0 & -1 & 0 & Q_2 = (Q_2^1,\ldots,Q_2^{16}) & 0 & 1 & 0 
\\ \hline 
E_{3} & \sfrac 12 & \sfrac 12  & 0 & 0 & 0 & -1 & Q_3 = (Q_3^1,\ldots,Q_3^{16}) & 0 & 0 & 1 
\\ \hline 
}
\]
\caption{ \label{tb:C3Z22glsm} 
Superfield charge table for resolutions of the non--compact $\Cplx^3/\Intr_2\times\Intr_2$ orbifold.}
\end{table}

This section focus on heterotic resolutions of the non--compact $\Cplx^3/\Intr_2\times\Intr_2$ using (0,2) GLSMs. (Some ingredients of the present discussion are inspired by ref.~\cite{GrootNibbelink:2010qut}.) The three complex coordinates $z_u$, $u=1,2,3$, of $\Cplx^3$ augmented with three exceptional coordinates $x_r$, $r=1,2,3$, to describe the resolution. These coordinates become part of the chiral superfields $\gF_u$ and $\gF_r'$ on which three $U(1)$ gauge symmetries $E_r$ act according to the charge table~\ref{tb:C3Z22glsm}. In this table the unit charged chiral superfields $\gO_r$ are composite, {\em i.e.}\ functions of the fundamental superfields $\gF_u$ and $\gF_r'$.

\subsection{Geometrical interpretation}

The analysis of the geometrical interpretation of this GLSM starts with writing down the D--term equations
\sequ{eq:DtermsC3Z22}{ 
\frac{1}{2}\left|z_{2}\right|^{2}+\frac{1}{2}\left|z_{3}\right|^{2}=b_{1}+\left|x_{1}\right|^{2}~, 
\\[1ex] 
\frac{1}{2}\left|z_{1}\right|^{2}+\frac{1}{2}\left|z_{3}\right|^{2}=b_{2}+\left|x_{2}\right|^{2} ~,
\\[1ex] 
\frac{1}{2}\left|z_{1}\right|^{2}+\frac{1}{2}\left|z_{2}\right|^{2}=b_{3}+\left|x_{3}\right|^{2}~. 
} 
Here the three parameters $b_r$ are the real parts of the three FI--parameters $\gr_r$ associated with the three gaugings $E_r$ which are assumed to be constant. An equivalent but useful representation of these equations are obtained by adding two of them and subtracting the third: 
\sequ{eq:DtermsC3Z22mod}{  \label{eq:DtermsC3Z22moda} 
\left|z_{1}\right|^{2}+\left|x_{1}\right|^{2}=b_{2}+b_{3}-b_{1}+\left|x_{2}\right|^{2}+\left|x_{3}\right|^{2}~, 
\\[2ex] \label{eq:DtermsC3Z22modb} 
\left|z_{2}\right|^{2}+\left|x_{2}\right|^{2}=b_{1}+b_{3}-b_{2}+\left|x_{1}\right|^{2}+\left|x_{3}\right|^{2}~,
\\[2ex] \label{eq:DtermsC3Z22modc} 
\left|z_{3}\right|^{2}+\left|x_{3}\right|^{2}=b_{1}+b_{2}-b_{3}+\left|x_{1}\right|^{2}+\left|x_{2}\right|^{2}~.
}
Depending on the relative values of the three FI--parameters the model can be in a number of phases which have different geometrical interpretations~\cite{GrootNibbelink:2010qut}. Here not all of them are listed and discussed, instead, the focus is on a number of particular interesting phases: the orbifold phase and the three full resolved phases which are characterised by having all three FI--parameters negative or positive, respectively. Other phases, in which some FI--parameters are positive while others are negative, correspond to partial blowups and are ignored here. (In ref.~\cite{GrootNibbelink:2010qut} some aspects of these other phases were investigated.)

Some topological properties of the effective geometries in the various phases can be determined. The divisors in the effective geometry can be identified by setting one of the complex coordinates to zero while satisfying all the D--term equations. The ordinary divisors are defined by $D_u := \{z_u = 0\}$ and the exceptional ones by $E_r := \{x_r = 0\}$.  The results of this analysis are summarised in Table~\ref{tb:C3Z22Phases}.

For each set of non--vanishing fields $Z_{(P)}=(Z_{(P)}^1,Z_{(P)}^2,Z_{(P)}^3)$, that defines a coordinate patch within a phase of the resolution GLSM, the other the complement set of fields $\{\tZ_{(P)}^1,\tZ_{(P)}^2,\tZ_{(P)}^3\}\in\Real^3$ then define a coordinate patch. The resulting patches are also given in Table~\ref{tb:C3Z22Phases}. A gauge can be chosen such that the phases of these non--zero fields $Z_{(P)}$ are all trivial, i.e.\ multiplets of $2\pi i$. This only leaves residual discrete gauge transformations in each of these patches: 
\equ{
Z_{(P)}^a \ra e^{i (\cQ_{(P)})^a{}_r\, \ga^p} \, Z_{(P)}^a \stackrel{!}{=} e^{2\pi i\, m^a}\, Z_{(P)}^a~, 
}
where $Z_{(P)}^a$, $a=1,2,3$, are the three scalar fields that do not vanish in patch $(P)$ with charges $ (\cQ_{(P)})^a{}_r$ and $m^a$ are integers. For the coordinate patches under investigation the charge matrices are given in~\eqref{eq:ChargeMatricesC3Z22}. Hence, the gauge parameters of the residual gauge transformations read: 
\equ{
\ga^T = 2\pi\, m^T\, \cQ_{(P)}^{-T}~. 
}
with $\ga = \big(\ga_1, \ga_2, \ga_3\big)$ and $m^T = \big(m^1, m^2, m^3\big)$. 
This induces residual gauge transformation on the coordinates of the coordinate patch $(P)$ transform 
\equ{ \label{eq:ResidualGaugeTransformationC3Z22} 
\tZ_{(P)}^a \ra e^{i (\widetilde{\cQ}_{(P)})^a{}_r\, \ga^p} \, \tZ_{(P)}^a 
= e^{2\pi i\, (\cR m)^a}\,\tZ_{(P)}^a~, 
\qquad 
\cR_{(P)} = \widetilde{\cQ}_{(P)}\cQ_{(P)}^{-1}
}
where $\tQ_{(P)}$ are the charges of the coordinates of the patch which are given in~\eqref{eq:ChargeMatricesC3Z22conjugate}. Thus if $\cR_{(P)}$ is integral, the residual gauge transformations are trivial.

\begin{table} 
\begin{center}
\begin{tabular}{|c||c|c|c|c|} 
 \hline
Phase & Non--zero fields & Patches & Curves & Intersection 
\\ \hline\hline
   Orbifold & $x_1,x_2,x_3\neq0$ & $(O) := \{z_{1},z_{2},z_{3}\}$  & $D_1D_2$, $D_2D_3$, $D_3D_1$ & $D_{1}D_{2}D_{3}$ 
\\  \hline\hline 
S--triangulation & $z_1,z_2,z_3\neq0$ &  $(S):=\{x_1,x_2,x_3\}$ & $E_1E_2$, $E_2E_3$, $E_3E_1$ & $E_{1}E_{2}E_{3}$ \\ \cline{2-5} 
  & $z_1,z_2,x_3\neq0$ & $(33):=\{x_1,x_2,z_3\}$ & $E_1E_2$, $E_2D_3$, $D_3E_1$ & $E_{1}E_{2}D_{3}$  \\  \cline{2-5} 
  & $z_1,x_2,z_3\neq0$ & $(22):=\{x_1,x_3,z_2\}$ & $E_1E_3$, $E_1D_2$, $D_2E_3$  &$E_{1}E_{3}D_{2}$  \\  \cline{2-5} 
  & $x_1,z_2,z_3\neq0$ & $(11):=\{x_2,x_3,z_1\}$ & $E_2E_3$, $E_2D_1$, $D_1E_3$ & $E_{2}E_{3}D_{1}$  
\\ \hline\hline
   E$_{1}$--triangulation 
   & $z_{2},z_{3},x_{3}\neq 0$ & $(31):=\{x_1,x_2,z_1\}$ & $E_1E_2$, $E_2D_1$, $D_1E_1$ & $E_{1}E_{2}D_{1}$  \\  \cline{2-5} 
   & $z_{1},z_{2},x_{3}\neq 0$ & $(33):=\{x_1,x_2,z_3\}$ & $E_1E_2$, $E_2D_3$, $D_3E_1$ & $E_{1}E_{2}D_{3}$ \\  \cline{2-5} 
   & $z_{2},z_{3},x_{2}\neq 0$ & $(21):=\{x_1,x_3,z_1\}$ & $E_1E_3$, $E_3D_1$, $D_1E_1$ & $E_{1}E_{3}D_{1}$ \\  \cline{2-5} 
   & $z_{1},z_{3},x_{2}\neq 0$ & $(22):=\{x_1,x_3,z_2\}$ & $E_1E_3$, $E_3D_2$, $D_2E_1$ & $E_{1}E_{3}D_{2}$ 
 \\ \hline\hline
   E$_{2}$--triangulation 
   & $z_{1},z_{3},x_{3}\neq 0$ & $(32):=\{x_1,x_2,z_2\}$ & $E_1E_2$, $E_2D_2$, $D_2E_1$ & $E_{1}E_{2}D_{2}$  \\  \cline{2-5} 
   & $z_{1},z_{2},x_{3}\neq 0$ & $(33):=\{x_1,x_2,z_3\}$ & $E_1E_2$, $E_2D_3$, $D_3E_1$ & $E_{1}E_{2}D_{3}$ \\  \cline{2-5} 
   & $z_{2},z_{3},x_{1}\neq 0$ & $(11):=\{x_2,x_3,z_1\}$ & $E_2E_3$, $E_3D_1$, $D_1E_2$ & $E_{2}E_{3}D_{1}$ \\  \cline{2-5} 
   & $z_{1},z_{3},x_{1}\neq 0$ & $(12):=\{x_2,x_3,z_2\}$ & $E_2E_3$, $E_3D_2$, $D_2E_2$ & $E_{2}E_{3}D_{2}$ 
 \\ \hline\hline
   E$_{3}$--triangulation 
   & $z_{2},z_{3},x_{1}\neq 0$ & $(11):=\{x_2,x_3,z_1\}$ & $E_2E_3$, $E_3D_1$, $D_1E_2$ & $E_{2}E_{3}D_{1}$  \\  \cline{2-5} 
   & $z_{1},z_{2},x_{1}\neq 0$ & $(13):=\{x_2,x_3,z_3\}$ & $E_2E_3$, $E_3D_3$, $D_3E_2$ & $E_{2}E_{3}D_{3}$ \\  \cline{2-5} 
   & $z_{1},z_{2},x_{2}\neq 0$ & $(23):=\{x_1,x_3,z_3\}$ & $E_1E_3$, $E_3D_3$, $D_3E_1$ & $E_{1}E_{3}D_{3}$ \\  \cline{2-5} 
   & $z_{1},z_{3},x_{2}\neq 0$ & $(22):=\{x_1,x_3,z_2\}$ & $E_1E_3$, $E_3D_2$, $D_2E_1$ & $E_{1}E_{3}D_{2}$ 
 \\ \hline
\end{tabular}
\end{center} 
\caption{\label{tb:C3Z22Phases} 
This table indicates which combination of fields are necessarily non--vanishing in  the orbifold and the three full resolution phases. This in turn determines the coordinate patches of the phases and hence the curves and intersections that exist within the patches. 
The notation $(ru)$ of the patches of the fully resolved geometries signify that the coordinates $x_r$ and $z_{v\neq u}$ are non--zero.  }
\end{table}

\subsubsection*{Orbifold phase} 

In the orbifold regime all three K\"ahler parameters are negative: $b_{1}, b_{2}, b_{3}<0$. The D--term equations~\eqref{eq:DtermsC3Z22} then imply that all three exceptional coordinates are non--vanishing: 
\sequ{eq:OrbiPhaseC3Z22}{ 
\left|x_{1}\right|^{2}=-b_{1}+\left|z_{2}\right|^{2}+\left|z_{3}\right|^{2}>0~,
\\[2ex] 
\left|x_{2}\right|^{2}=-b_{2}+\left|z_{1}\right|^{2}+\left|z_{3}\right|^{2}>0~,
\\[2ex] 
\left|x_{3}\right|^{2}=-b_{3}+\left|z_{1}\right|^{2}+\left|z_{2}\right|^{2}>0~,
}
hence there is a single coordinate patch: $\{z_{1}, z_{2}, z_{3}\}$. In particular, the D--term equations allow to set all these three coordinates to zero at the same time. Moreover, it is clear that none of the exceptional divisors $E_r$ exist in this phase. Instead the intersection of $D_1D_2D_3$ exists.

By exploiting the gauge symmetries it is possible to fix the phases of $x_1, x_2, x_3$ some arbitrary values (which are typically taken to be zero for simplicity). However, these gauge fixings do not fix the gauges completely, since the matrix~\eqref{eq:ResidualGaugeTransformationC3Z22} in this case, 
\equ{
\cR_{(O)} = \widetilde{\cQ}_{(O)}\cQ_{(O)}^{-1} = -\widetilde{\cQ}_{(O)}  
= \pmtrx{
~0~ & ~\sfrac 12~ & ~\sfrac 12~ \\ 
\sfrac 12 & 0 & \sfrac 12 \\
\sfrac 12 & \sfrac 12 & 0 
}~, 
}
is non--integer, therefore, there are non--trivial residual $\Intr_2$ gauge transformations which act as 
\sequ{eq:ResidualOrbiC3Z22GaugeTrans}{
E_1~:~ (z_1, z_2, z_3) \ra (z_1, -z_2, -z_3)~, 
\\[2ex]
E_2~:~ (z_1, z_2, z_3) \ra (-z_1, z_2, -z_3)~, 
\\[2ex] 
E_3~:~ (z_1, z_2, z_3) \ra (-z_1, -z_2, z_3)\phantom{~,} 
}
on the remaining coordinates. The first two are precisely the transformations that defined the $\Cplx^3/\Intr_2\times \Intr_2$ orbifold and the third one is simply the combination of the first two and hence redundant in the orbifold phase.

\subsubsection*{S--triangulation full resolution phase}

In the S--triangulation the K\"ahler parameters satisfy the following inequalities:
\equ{ 
0<b_{3}<b_{1}+b_{2}~,
\quad 
0<b_{2}<b_{1}+b_{3}~, 
\quad 
0<b_{1}<b_{2}+b_{3}~. 
}
From \eqref{eq:DtermsC3Z22} it follows that at least two of the three $z_{u}$ are non--zero. Hence, there is one coordinate patch $\{x_{1}, x_{2}, x_{3}\}$ when all three $z_u$ are non--vanishing. Taking \eqref{eq:DtermsC3Z22mod} into account, there are, in addition, three coordinate patches $\{z_{u}, x_{p \neq u}\}$ for $u=1,2,3$ when $x_u$ and $z_{p\neq u}$ are non--zero. 

There is no non--trivial residual gauge transformation on the coordinate patch $(S):=\{x_{1}, x_{2},x_{3}\}$, since fixing the phases of all three $z_u$ fixes all gauge parameters $\gth_r$ up to multiples of $2\pi i$, hence the actions on the coordinates $x_r$ are trivial. For the coordinate patch $(33):=\{z_1, x_2,x_3\}$ the non--vanishing coordinates of which the phases can be set to unity are $x_1, z_2, z_3$, consequently, the gauge parameters $\gth_{2,3}$ are fixed modulo multiples of $4\pi i$ and $\gth_1$ modulo multiplets of $2\pi i$. But the residual gauge transformations on coordinates 
\equ{ 
z_1 \ra e^{\sfrac 12\, \gth_2+\sfrac 12\, \gth_3} z_1~,
\quad 
x_2 \ra e^{-\gth_2} x_2~, 
\quad 
x_3 \ra e^{-\gth_3} x_3
}
in the patch $(33)$ only involve the gauge parameters $\gth_{2,3}$, and hence these phase transformations are trivial. Similar arguments can be provided for the other patches $(22):=\{x_1,x_3,z_2\}$ and $(11):=\{x_2,x_3,z_1\}$. The fact that all the coordinate patches of this triangulation are regular can also be verified by showing that the matrices $\cR_{(P)}$ defined in~\eqref{eq:ResidualGaugeTransformationC3Z22} are all integral.

It follows that in the S--triangulation all the divisors $D_u$ and $E_r$ exist, though not in all coordinate patches. Aside from the curves $E_r D_{u\neq r}$, all three exceptional curves $E_1E_2$, $E_2E_3$ and $E_3E_1$ exist. In particular, the intersections 
\equ{ 
E_{1} E_{2} E_{3}=E_{2} E_{3} D_{1}=E_{1} E_{3} D_{2}=E_{1} E_{2} D_{3} = 1
}
are all equal to unity as there is just a single solution to the D--term equations and there is no residual gauge transformation acting on the coordinates in any given coordinate patch. All this information is encoded in the toric diagram for the S--triangulation: 
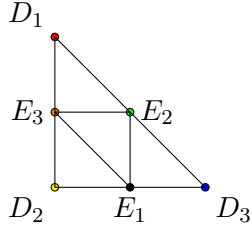
\begin{center}
\begin{tikzpicture}
\foreach \Point/\PointLabel in { (-4,-1)/D_ {2}}
\draw[fill=yellow] \Point circle (0.05) node[below left]
{$\PointLabel$};
\foreach \Point/\PointLabel in { (-4,0)/E_ {3}}
\draw[fill=orange] \Point circle (0.05) node[left]
{$\PointLabel$};
\foreach \Point/\PointLabel in { (-4,1)/D_ {1}}
\draw[fill=red] \Point circle (0.05) node[above left]
{$\PointLabel$};
\foreach \Point/\PointLabel in { (-3,0)/E_ {2}}
\draw[fill=green] \Point circle (0.05) node[right]
{$\PointLabel$};
\foreach \Point/\PointLabel in { (-2,-1)/D_ {3}}
\draw[fill=blue] \Point circle (0.05) node[below right]
{$\PointLabel$};
\foreach \Point/\PointLabel in { (-3,-1)/E_ {1}}
\draw[fill=black] \Point circle (0.05) node[below]
{$\PointLabel$};
\draw (-3,-1) -- (-4,0); 
\draw (-3,-1) -- (-3,0); 
\draw (-4,0) -- (-3,0); 
\draw (-4,-1) -- (-2,-1)  -- (-4,1) -- (-4,-1);
\underline{A variation annomaly field redefinition}
\end{tikzpicture}
\end{center}
Indeed, all the divisors are indicated as dots. The existing curves are represented as lines between two adjacent dots and the unit intersections are the smallest triangles in the diagram. At the same time these smallest triangles also indicate the four coordinate patches.

\subsubsection*{E$_1$--triangulation full resolution phase}

In the E$_{1}$--triangulation the K\"ahler parameters satisfy the conditions
\equ{ 
 0<b_{2}+b_{3}<b_{1}~,
 \quad
 0<b_{3}<b_{1}+b_{2}~,
 \quad 
 0<b_{2}<b_{1}+b_{3}~. 
}  
Again at least two of the three $z_u$ are non--zero. In light of the first inequality above, it is convenient to write the equation \eqref{eq:DtermsC3Z22moda} as 
\equ{
\left|x_{2}\right|^{2}+\left|x_{3}\right|^{2}
=b_1 -b_{2}-b_{3}+ 
\left|z_{1}\right|^{2}+\left|x_{1}\right|^{2}~.  
} 
Hence either $x_2$ or $x_3$ is non--zero. If $x_2\neq 0$ then \eqref{eq:DtermsC3Z22modc} implies that $z_3$ is non--vanish as there needs to at least two $z_u\neq0$. Similarly, if $x_3\neq0$ then  \eqref{eq:DtermsC3Z22modb} says that $z_2$ is non--vanishing. Therefore, in total there are four coordinate patches:  $\{x_1,x_2,z_1\}$,  $\{x_1,x_2,z_3\}$, $\{x_1,x_3,z_1\}$, $\{x_1,x_3,z_2\}$. Again all these patches are regular; there is no residual orbifold action on them. 

In this phase the exceptional curves $E_1E_2$ and $E_1E_3$ exist but $E_2E_3$ does not. Instead the curve $D_1E_1$ is allowed by the D--term equations. The following intersections
\equ{
E_{1} E_{2} D_{3}=E_{1} E_{3} D_{2}=E_{1} E_{2} D_{1}=E_{1} E_{3} D_{1}=1
} 
are all equal to unity. All this information is encoded in the toric diagram for the E$_1$--triangulation: 
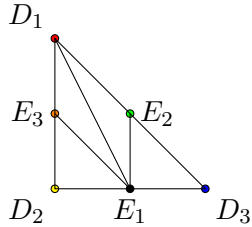
\begin{center}
\begin{tikzpicture}
\foreach \Point/\PointLabel in { (-4,3)/D_ {2}}
\draw[fill=yellow] \Point circle (0.05) node[below left]
{$\PointLabel$};
\foreach \Point/\PointLabel in { (-4,4)/E_ {3}}
\draw[fill=orange] \Point circle (0.05) node[left]
{$\PointLabel$};
\foreach \Point/\PointLabel in { (-4,5)/D_ {1}}
\draw[fill=red] \Point circle (0.05) node[above left]
{$\PointLabel$};
\foreach \Point/\PointLabel in { (-3,4)/E_ {2}}
\draw[fill=green] \Point circle (0.05) node[right]
{$\PointLabel$};
\foreach \Point/\PointLabel in { (-2,3)/D_ {3}}
\draw[fill=blue] \Point circle (0.05) node[below right]
{$\PointLabel$};
\foreach \Point/\PointLabel in { (-3,3)/E_ {1}}
\draw[fill=black] \Point circle (0.05) node[below]
{$\PointLabel$};
\draw (-4,4) -- (-3,3); 
\draw (-3,4) -- (-3,3); 
\draw (-4,5) -- (-3,3); 
\draw (-4,3) -- (-2,3)  -- (-4,5) -- (-4,3);
\end{tikzpicture}
\end{center}

A similar analysis can be performed for the other two full resolution phases corresponding to the triangulations  E$_{2}$ and E$_{3}$. A summary of the results are given in table~\ref{tb:C3Z22Phases}.

\subsection{Pairs of GLSMs associated to torsion related orbifolds}
\label{sc:PairsTorsionGLSMs}

The charges of the Fermi superfields are kept arbitrary in table~\ref{tb:C3Z22glsm}. In order that the GLSM is free of gauge anomalies these charge vectors are subject to the conditions~\cite{GrootNibbelink:2010qut}
\equ{ \label{eq:AnomalyCancellationC3Z22}
 Q_{1}^{2}=Q_{2}^{2}=Q_{3}^{2}=\frac{3}{2}~, 
 \qquad 
Q_{1} \cdot Q_{2}=Q_{2} \cdot Q_{3}=Q_{3} \cdot Q_{1}=\frac{1}{4} 
}
and the sum of charges for each of the three gaugings vanishes, see Subsection~\ref{sc:AnomalyConsistencyConditions}. The first three equations indicates that consistent choices for the charge vectors are given by the shifted momenta of the three twisted sectors without oscillators, see table~\ref{tb:TwistedSpectrumWithOutOscillators}, since they all square to $3/2$. The latter three equations can be satisfied by taking the shifted momenta 
\equ{
\hspace{-1ex} 
Q_1 = \big(0, \sfrac{1}{2},\sfrac{1}{2}, \sm1,0,0,0^{2}\big)\big(0^{8}\big)~;~~
Q_2 = \big(\sfrac{1}{2},0,\sfrac{1}{2},0,\sm1,0,0^{2}\big)\big(0^{8}\big)~;~~
Q_3 = \big(\sfrac{1}{2},\sfrac{1}{2},0,0,0,\sm1,0^{2}\big)\big(0^{8}\big) 
}
out of the three twisted sectors of the orbifold model without discrete torsion or by 
\equ{
\hspace{-1ex} 
Q_1^{\!\times} = \sm\big(0, \sfrac{1}{2},\sfrac{1}{2}, \sm1,0,0,0^{2}\big)\big(0^{8}\big);\,
Q_2^{\!\times} = \sm\big(\sfrac{1}{2},0,\sfrac{1}{2},0,\sm1,0,0^{2}\big)\big(0^{8}\big);\,  
Q_3^{\!\times} = \sm\big(\sfrac{1}{2},\sfrac{1}{2},0,0,0,\sm1,0^{2}\big)\big(0^{8}\big) 
}
of the orbifold model with torsion. Notice that this is precisely how the brother gauge shift vectors were related to the original ones as discussed in Subsection~\ref{sc:DiscreteTorsionPhase}.

These certainly do not represent unique choices, but for any choice of anomaly free charge vectors from shifted momenta of the physical twisted states without oscillators in the orbifold model without torsion, the choice of the same charge vectors but all with the opposite sign, is an anomaly free choice with torsion. Hence, switching torsion on or off corresponds to the mapping
\equ{ 
Q_1 \leftrightarrow Q_1^{\!\times} = -Q_1~, 
\qquad
Q_2 \leftrightarrow Q_2^{\!\times} = -Q_2~, 
\qquad
Q_3 \leftrightarrow Q_3^{\!\times} = -Q_3
}
of all the charges in the two associated resolution GLSMs simultaneous. This suggests that there is a field redefinition from the Fermi superfields $\gL$ in the non--torsion model to the Fermi superfields $\gL^{\!\!\times}$ in the torsion model. Formally, in terms of the chiral superfields $\gO_r$ defined in table~\ref{tb:C3Z22glsm} this superfield redefinition can be stated as
\equ{\label{eq:SuperfieldRedefinition}
\gL \ra 
\gL^{\!\!\times} = e^{-2\log \gO\cdot Q}\, \gL~, 
}
since this precisely reverses all the charges of $\gL$. In order that this field redefinition is well--defined $\gO_r$ should be non--singular. Given that in various coordinate patches within the phases of the theory, there are always three superfields non--vanishing they can be used in this field redefinition. Table~\ref{tb:OmegaRepresentationsC3Z22Phases} summarises the choices for $\gO_r$ in the patches under investigation here.

\begin{table} 
\begin{center}
\begin{tabular}{|c||c|c|c|c|} 
 \hline
Phase & Patch & \multicolumn{3}{c|}{Non--singular superfield representation of} 
\\
&(P) &${\Omega_{1}}$&${\Omega_{2}}$&${\Omega_{3}}$
\\ \hline\hline
Orbifold & $(O)$  & \quad\qquad $\gF_{1}^{\prime-1}$\qquad\quad  & \quad\qquad$\gF_{2}^{\prime-1}$\quad\qquad & \quad\qquad$\gF_{3}^{\prime-1}$\qquad\quad 
\\ \hline\hline
S--triangulation & $(S)$ & $\gF_{2}\gF_{3}\gF_{1}^{-1}$ & $\gF_{1}\gF_{3}\gF_{2}^{-1}$  &$\gF_{1}\gF_{2}\gF_{3}^{-1}$ 
\\ \cline{2-5}
 & $(33)$ & $\gF_{2}^{2}\gF_{3}^\prime$ & $\gF_{1}^{2}\gF_{3}^\prime$ & $\gF_{3}^{\prime-1}$
\\ \cline{2-5}
 & $(22)$ & $\gF_{3}^{2}\gF_{2}^\prime$ & $\gF_{2}^{\prime-1}$ & $\gF_{1}^{2}\gF_{2}^\prime$
\\ \cline{2-5}
 & $(11)$ & $\gF_{1}^{\prime-1}$ & $\gF_{3}^{2}\gF_{1}^\prime$ & $\gF_{2}^{2}\gF_{1}^\prime$
\\ \hline\hline
E$_{1}$--triangulation & $(31)$ & $\gF_{2}^{2}\gF_{3}^\prime$ & $\gF_{3}^{2}\gF_{2}^{-2}\gF_{3}^{\prime-1}$ & $\gF_{3}^{\prime-1}$
\\ \cline{2-5}
& $(33)$ & $\gF_{2}^{2}\gF_{3}^\prime$ & $\gF_{1}^{2}\gF_{3}^\prime$  &  $\gF_{3}^{\prime-1}$
\\ \cline{2-5}
& $(21)$ &  $\gF_{3}^{2}\gF_{2}^\prime$& $\gF_{2}^{\prime-1}$ & $\gF_{2}^{2}\gF_{3}^{-2}\gF_{2}^{\prime-1}$
\\ \cline{2-5}
& $(22)$ & $\gF_{3}^{2}\gF_{2}^\prime$ &  $\gF_{2}^{\prime-1}$ &  $\gF_{1}^{2}\gF_{2}^\prime$
\\ \hline\hline 
E$_{2}$--triangulation  & $(32)$ & $\gF_1^{-2}\gF_3^{2}\gF_3^{\prime-1}$ & $\gF_1^2\gF_3^\prime$ & $\gF_3^{\prime-1}$ 
 \\ \cline{2-5} 
& $(33)$ & $\gF_{2}^{2}\gF_{3}^\prime$ & $\gF_{1}^{2}\gF_{3}^\prime$ & $\gF_{3}^{\prime-1}$
\\ \cline{2-5} 
 & $(11)$ & $\gF_{1}^{\prime-1}$ & $\gF_{3}^{2}\gF_{1}^\prime$ & $\gF_{2}^{2}\gF_{1}^\prime$ 
\\  \cline{2-5} 
& $(12)$ & $\gF_1^{\prime-1}$ & $\gF_3^2\gF_1^\prime$ & $\gF_1^2 \gF_3^{-2} \gF_1^{\prime-1}$
 \\ \hline\hline
E$_{3}$--triangulation & $(11)$ & $\gF_{1}^{\prime-1}$ & $\gF_{3}^{2}\gF_{1}^\prime$ & $\gF_{2}^{2}\gF_{1}^\prime$
\\  \cline{2-5} 
& $(13)$ & $\gF_1^{\prime-1}$ & $\gF_2^2\gF_3^{-2}\gF_1^{\prime-1}$ & $\gF_3^2\gF_1^\prime$  
\\ \cline{2-5} 
& $(23)$ & $\gF_1^{-2}\gF_2^2\gF_2^{\prime-1}$ & $\gF_2^{\prime-1}$ & $\gF_1^2\gF_2^\prime$
\\  \cline{2-5} 
& $(22)$ & $\gF_{3}^{2}\gF_{2}^\prime$ & $\gF_{2}^{\prime-1}$ & $\gF_{1}^{2}\gF_{2}^\prime$
\\ \hline
\end{tabular}
\end{center}
\caption{\label{tb:OmegaRepresentationsC3Z22Phases} 
This table gives the explicite non--singular forms of $\gO_r$ the orbifold and the full resolution patches in the three triangulations. }
\end{table}

Notice that~\eqref{eq:SuperfieldRedefinition} precisely looks like a super gauge transformation~\eqref{eq:GaugeTransFermiMultiplet} but with the super gauge parameters $\gTh$ replaced by $-2\log \gO$. Since only the Fermi multiplet are involved in this superfield redefinition, it is anomalous. Because this superfield redefinition is of the same form as a super gauge transformation, the form of the anomaly is known to be 
\equ{ \label{eq:AnomSuperfieldRedefinition} 
W_\text{sf\, redef\, anom} = 
-\frac 1{2\gp} \,   \sum_{r, s} \cA_{rs}\, \log \gO^{r} F^{s}
= -\frac 1{2\gp} \, 
\Big\{ 
 \frac 32 \sum_r  \log \gO_r \, F^r 
+  \frac 14 \sum_{s\neq r} \log \gO_{s} \, F^{r} 
\Big\}~, 
}
using the general form of the super gauge anomaly~\eqref{eq:SuperGaugeAnom}. The latter form is obtained by using the explicit expression~\eqref{eq:AnomalyCancellationC3Z22} of the anomaly matrix $\cA_{rs} = Q_r\cdot Q_s$ given by 
\equ{ \label{eq:AnomMatrixSuperfieldRedefinition}
\cA = \pmtrx{
~\sfrac 32~ & ~\sfrac 14~ & ~\sfrac 14~ 
\\
 \sfrac 14  &\sfrac 32 & \sfrac 14 
 \\  
 \sfrac 14 & \sfrac 14 &  \sfrac 32 
}~. 
}
The superfield anomaly~\eqref{eq:AnomSuperfieldRedefinition} is of the form of superfield dependent FI--actions~\eqref{eq:GLSM_FI} but with the FI--parameters $\gr_r$ replaced by 
\equ{
\gr^T \ra \gr^{\!\times \,T} =  \gr^T - \frac 1{2\gp}\, \log \gO^T\, \cA~.  
}
where $\gr^T = \big(\gr_1, \gr_2, \gr_3\big)$ and $\log \gO^T = \big(\log \gO_1, \log \gO_2, \log \gO_3\big)$.

The field redefinition anomaly~\eqref{eq:AnomSuperfieldRedefinition} is not gauge invariant: it gives a phase in the Euclidean path integral 
\equ{
\gd_\gTh S_\text{sf\,redef\,anom} \supset -i \int \ga_r\, \cA_{rs}\ \frac{F^s_{E2}}{2\pi}~. 
}
However, since it is only obtained under the assumption that the field redefinition~\eqref{eq:SuperfieldRedefinition} is non--singular, it only receives discrete phase contributions from the scalar fields in Table~\ref{tb:C3Z22Phases} that do not vanish.

The flux quantisation conditions~\eqref{eq:FluxQuantisation} for the present GLSM read 
\sequ{}{ 
\int \frac{F_{E_{2}}^{1}}{2\pi} \in \mathbb{Z},
\qquad 
\int \frac{F_{E_{2}}^{2}}{2\pi} \in \mathbb{Z},
\qquad
\int \frac{F_{E_{2}}^{3}}{2\pi} \in \mathbb{Z}~,
\\[2ex] 
\frac 12\, \int \frac{F_{E_{2}}^{2}}{2\pi}+\frac 12\, \int \frac{F_{E_{2}}^{3}}{2\pi} \in \mathbb{Z},
\quad
\frac 12\, \int \frac{F_{E_{2}}^{1}}{2\pi}+\frac 12\,  \int \frac{F_{E_{2}}^{3}}{2\pi} \in \mathbb{Z},
\quad
\frac 12\, \int \frac{F_{E_{2}}^{1}}{2\pi}+\frac 12\, \int \frac{F_{E_{2}}^{2}}{2\pi} \in \mathbb{Z}~.
}
The first three conditions follow from the charges of the chiral superfields $\gF_r'$ and the latter three from those of $\gF_u$. Thus all gauge fluxes are integers and the sums of two gauge fluxes are even integers. The latter quantisation conditions are solved by adding two equations and subtracting the third: 
\equ{ 
\frac 1{2\gp} \int \pmtrx{ F_{E2}^1 \\ F_{E2}^2 \\ F_{E2}^3 } 
= \cF\, n~, 
\qquad 
\cF = \cQ_{(S)}^{-1}~, 
}
in terms of three integers $n^T=\big(n_1,n_2,n_3\big)$. Here, $\cQ_{(S)}$ is one of the charge matrices defined in~\eqref{eq:ChargeMatricesC3Z22}
of Appendix~\ref{sc:ChargeMatrices} and their inverse transposed forms in~\eqref{eq:ChargeMatricesC3Z22TransposedInverse}. As can be seen from there, $\cF$ is an integral matrix, the first three quantisation conditions are fulfilled as well. As was argued in~\cite{Adams:2009tt} possible vacuum phases in (orbifold) partition functions may be recovered in the GLSM as non--invariances of the path integral encoded in 
\equ{
\gd_\gTh S_\text{sf\, redef\, anom} \supset -2\gp i \, m^T \cM_{(P)}\, n~~, 
\qquad 
\cM_{(P)} = \cQ_r^{-T} \cA \cF = \cQ_{(P)}^{-T} \cA\cQ_{(S)}^{-1}~. 
}
Hence, the path integral is invariant if $\cM_{(P)}$ is an integral matrix. By explicit matrix multiplications it may be confirmed that $\cM_{(P)}$ is indeed integral for all charge matrices~\eqref{eq:ChargeMatricesC3Z22} that correspond to any of the patches of the three fully resolved phases. On the contrary in the orbifold phase one finds:
\equ{
\cM_{(O)} = \pmtrx{
~1~ & \,\sm\sfrac 32~ & \,\sm\sfrac 32~ \\
\,\sm\sfrac 32~ & ~1~ & \,\sm\sfrac 32~ \\ 
\,\sm\sfrac 32~ & \,\sm\sfrac 32~ & ~1~ 
} 
\equiv \pmtrx{
0 & \sfrac 12 & \sfrac 12 \\ 
\sfrac 12 & 0 & \sfrac 12 \\ 
\sfrac 12 & \sfrac 12 & 0 
}~. 
}
The final expression is obtained modulo integral matrices. This shows that in the orbifold phase the discrete torsion phases are reproduced by the residual gauge transformations of the field redefinition anomaly.

To summarise, the two non--compact resolution GLSMs associated to the orbifold theories with and without torsions are both free of gauge anomalies and hence consistent models. The effect of discrete torsion between the two models is recovered in their orbifold phases, if both models are expressed in the same field basis ({i.e.}\ with chiral Fermi multiplets with the same gauge charges in both models) because of a field redefinition anomaly~\eqref{eq:AnomSuperfieldRedefinition}. Even though in this expression there are logs of chiral superfields, these are non--singular, because the superfields which appear in the field redefinition~\eqref{eq:SuperfieldRedefinition} do not vanish in the patch where the particular field redefinition is defined (see Table~\ref{tb:OmegaRepresentationsC3Z22Phases}). In particular this does not signify that the geometry has torsion or should be augmented with NS5--branes, since in the unitary gauge the FI--terms are constants in each patch, hence the three--form flux~\eqref{eq:Hflux} vanishes.

\setcounter{equation}{0}
\section{GLSMs for resolutions of $\boldsymbol{T^6/\Intr_2\times\Intr_2}$}
\label{sc:T6Z22glsms}

The study of $(0,2)$ resolution GLSMs of the toroidal orbifold $T^6/\Intr_2\times\Intr_2$ is more involved than those for the non--compact orbifold $\Cplx^3/\Intr_2\times\Intr_2$ considered in the previous section. First of all, additional ingredients are needed to describe the geometry as the orbifold is compact. And partially because of this also the description of possible gauge backgrounds is more complicated. Only with these aspects understood, the consequences of discrete torsion in the underlying orbifold model can be properly investigated. Therefore, first Subsections~\ref{sc:ConstructionCompactZ22ResolutionGLSMs} to~\ref{sc:MinFullResGLSM} are used to develop a both accurate and manageable description of resolution GLSMs associated with the singular $T^6/\Intr_2\times\Intr_2$ geometry dubbed a minimal full resolution model. Subsection~\ref{sc:GaugeBackgroundMinFullRes} then gives the GLSM for a particular gauge background using the same blowup modes as in the non--compact model studied in the previous section. Finally, Subsection~\ref{sc:GaugeBackgroundMinFullResTorsion} the GLSM for the compact orbifold model with discrete torsion is studied.

\subsection[Construction of resolution GLSMs for compact ${\Intr_2\times\Intr_2}$ orbifolds]{Construction of resolution GLSMs for compact $\boldsymbol{\Intr_2\times\Intr_2}$ orbifolds}
\label{sc:ConstructionCompactZ22ResolutionGLSMs}

To construct GLSMs that describe resolutions of toroidal orbifold geometries, the following steps need to be taken~\cite{Blaszczyk:2011hs}: 
\begin{enumerate}
\item Give GLSM descriptions for each of the three underlying two--tori compatible with the orbifold symmetries.
\item Add so--called exceptional gaugings to introduce the orbifold actions and define the exceptional cycles.
\item Confirm that there is a regime where the GLSM description can be interpreted as the orbifold geometry under consideration. 
\item Determine the regimes in which the GLSM description can be interpreted as resolved geometries. 
\end{enumerate}
This program was discussed in~\cite{Blaszczyk:2011hs} for $(2,2)$ models, but these steps can equally well be executed in the $(0,2)$ language, which is used throughout this work.

It is important to realise that there are a number of different $T^6/\Intr_2\times\Intr_2$ orbifolds depending on their underlying six--torus lattice, see {\em e.g.}~\cite{Donagi2008,FRTV,Athanasopoulos:2016aws}. The construction here is aimed to resolve the particular one with Hodge numbers $(51,3)$. Moreover, one single orbifold geometry may be associated to many different GLSMs, even if the target space gauge configurations are not considered. The descriptions differ in the number of exceptional gaugings. Descriptions in which for all exceptional cycles of the resolved geometry there are exceptional gaugings, were dubbed maximal full resolution GLSMs in ref.~\cite{Blaszczyk:2011hs}. On the other end there are GLSMs descriptions with the least number of exceptional gaugings such that still the effective geometry in appropriate regimes corresponds to fully resolved orbifold resolutions. Such models were called minimal full resolution GLSMs. Between these two extremes there is a whole variety of GLSMs. Some of these models cannot describe fully resolved geometries; while others do~\cite{Blaszczyk:2011hs}. The focus in this paper is on full resolution GLSMs only. Such resolution GLSMs might possess many different phases. Only the orbifold phase and fully smooth resolution phases are investigated in this work in detail, while all kinds of interesting other phases will be ignored.

Maximal full resolution GLSMs are the most complete in the sense that all the K\"ahler parameters associated to the volumes of the exceptional cycles are made explicit. On the down side, this means that such models typically contain a large number of $U(1)$ gauge symmetries. As is discussed below the maximal full resolution GLSM for the toroidal orbifold $T^6/\Intr_2\times\Intr_2$ contains 51 $U(1)$ gaugings: for each of the 51 K\"ahler parameters there is a dedicated gauging available.  The minimal full resolution GLSM for this orbifold only requires six $U(1)$ gaugings: The radii of the three two--tori and collective volumes of the three types of exceptional cycles are explicit in that description.

Below, first the GLSMs description of a two--torus with $\Intr_2$ symmetries is recalled. After that the basic ingredients of the maximal full resolution GLSM are laid out. Details of the resulting geometry and the consequences of the discrete torsion of the orbifold model are investigated in the minimal full resolution GLSM only for simplicity.

\subsection[Two--tori GLSM with ${\Intr_2}$ symmetries]{Two--tori GLSM with $\boldsymbol{\Intr_2}$ symmetries}
\label{sc:TwoToriGLSM} 

\begin{table} 
\[
\arry{|c||cccc||cc|}{
\hline 
\text{Superfield} & \gF_{u\,1} & \gF_{u\,2} & \gF_{u\,3} & \gF_{u\,4} & \gG^{\phantom{\prime}}_u & \gG_u'  
\\ \hline 
\text{U(1) charges} & z_{u\,1} & z_{u\,2} & z_{u\,3} & z_{u\,4} & \gg^{\phantom{\prime}}_u & \gg_u'  
\\ \hline\hline 
R_{u'} & \sfrac 12\gd_{u'u} & \sfrac 12\gd_{u'u} & \sfrac 12\gd_{u'u} & \sfrac 12\gd_{u'u} & -\gd_{u'u} & -\gd_{u'u} 
\\ \hline 
}
\]
\caption{ \label{tb:T2wZ2glsm} 
Superfield charge table for the GLSM for three two--tori admiting $\Intr_2$ symmetries.}
\end{table}

In ref.~\cite{Blaszczyk:2011hs} it was argued that a convenient description of two--tori that admit $\Intr_2$ involutions are given by the superfields given in Table~\ref{tb:T2wZ2glsm} with the superpotential
\equ{ \label{eq:TwoTorusSuperpotential} 
P_\text{three two--tori} = \sum_u 
\Big( \gk_u\, \gF_{u\,1}^2 + \gF_{u\,2}^2 + \gF_{u\,3}^2\Big) \gG^{\phantom{\prime}}_u + 
\Big( \gF_{u\,1}^2 + \gF_{u\,2}^2 + \gF_{u\,4}^2\Big) \gG_u'~, 
}
where 
\equ{ \label{eq:ComplexStructureTwoTori} 
\gk_u = 
\frac{\cP_{\gt_u}(\sfrac{\gt_u}{2}) - \cP_{\gt_u}(\sfrac 12)}
{\cP_{\gt_u}(\sfrac{1+\gt_u}{2}) - \cP_{\gt_u}(\sfrac 12)} 
}
parameterise the complex structures $\gt_u$ of the three two--tori in terms of the Weierstrass $\cP$ function. This description was obtained as a rewriting of the well--known Weierstrass model of an eliptic curve. On each of the four chiral superfields  $\gF_{u\,x}$ of  two--torus $T^2_u$ a separate $\Intr_2$ reflection symmetry $\gF_{u\,x} \ra -\gF_{u\,x}$ can act leaving the superpotential invariant. In addition, there are two involutions per two--torus which can be identified with $\Intr_2$ translation on the two--torus lattice~\cite{Blaszczyk:2011hs}. The K\"ahler structures of the two--tori are encoded in the GLSM description as the FI--parameter $a_u$ associated with the gauging $R_u$. The resulting D-- and F--term equations in the conformal limit read: 
\begin{subequations}
\equ{
|z_{u\,1}|^2 + |z_{u\,2}|^2 + |z_{u\,3}|^2 + |z_{u\,4}|^2 = a_u~, 
\\[1ex] 
\gk_u\, z_{u\,1}^2 + z_{u\,2}^2 + z_{u\,3}^2 = 0~, 
\quad
z_{u\,1}^2 + z_{u\,2}^2 + z_{u\,4}^2 = 0~.  
}
\end{subequations} 
Because $\gk_u\neq 1$ the two F--term conditions can never combined to an equation with just two terms. Together with the $U(1)$ gaugings, which can remove a phase per $u$, shows that each set of $z_{u\,1},\ldots, z_{u\,4}$ coordinates for a given $u$ describes a geometry of real dimension two.

\subsection{Maximal full resolution GLSM} 

\begin{table} 
\[
\arry{|c||c|c|c||cc|cc|cc||c|c|c|}{
\hline 
\text{Superfield} & \gF_{1\,x} & \gF_{2\,y} & \gF_{3\,z} & \gG_1 & \gG_1' & \gG_2 & \gG_2' & \gG_3 & \gG_3' & \gF_{1\,yz}' & \gF_{2\,xz}' & \gF_{3\,xy}'  
\\ \hline 
\text{U(1) charge} & z_{1\,x} & z_{2\,y} & z_{3\,z} & \gg_1 & \gg_1' & \gg_2 & \gg_2' & \gg_3 & \gg_3' & x_{1\,yz} & x_{2\,xz} & x_{3\,xy}  
\\ \hline\hline 
R_1 & \sfrac 12 & 0 & 0 & -1 & -1 & 0 & 0 & 0 & 0 & 0 & 0 & 0 
\\ \hline 
R_2 & 0 & \sfrac 12 & 0 & 0 & 0 & -1 & -1 & 0 & 0 & 0 & 0 & 0 
\\ \hline 
R_3 & 0 & 0 & \sfrac 12 & 0 & 0 & 0 & 0 & -1 & -1 & 0 & 0 & 0 
\\ \hline\hline 
E_{1\,y'z'} & 0 & \sfrac 12 \gd_{y'y} & \sfrac 12 \gd_{z'z} & 0 & 0 & 0 & 0 & 0 & 0 & - \gd_{y'y}\gd_{z'z} & 0 & 0 
\\ \hline 
E_{2\,x'z'} & \sfrac 12\gd_{x'x} & 0 & \sfrac 12 \gd_{z'z} & 0 & 0 & 0 & 0 & 0 & 0 & 0 & -\gd_{x'x}\gd_{z'z} & 0 
\\ \hline 
E_{3\,x'y'} & \sfrac 12 \gd_{z'z} & \sfrac 12 \gd_{y'y} & 0 & 0 & 0 & 0 & 0 & 0 & 0 & 0 & 0 & -\gd_{x'x}\gd_{y'y} 
\\ \hline 
}
\]
\caption{ \label{tb:MaxT6Z22glsm} 
Superfield charge table that determines the geometry of maximal full resolution of $T^6/\Intr_2\times\Intr_2$.}
\end{table}

The maximal full resolution GLSM for the toroidal orbifold $T^6/\Intr_2\times\Intr_2$ has three ordinary gaugings $R_1$, $R_2$ and $R_3$ to define three two--tori and $3\cdot 16 = 48$ exceptional gaugings $E_{1,yz}$, $E_{2,xz}$ and $E_{3,xy}$ associated to the exceptional cycles. The full charge table is given in Table~\ref{tb:MaxT6Z22glsm}. The fermi superfields $\gG_1, \gG_1'$, $\gG_2, \gG_2'$ and $\gG_3,\gG_3'$ feature in the superpotential to define the three underlying two--tori, see~\eqref{eq:TwoTorusSuperpotential}. Because the exceptional gaugings the superpotential has to be extended to 
\equ{ \label{eq:T6Z22maxSuperpotential} 
\arry{rl}{
P_\text{max\,res}  & = 
\Big( 
\gk_1\, \gF_{1\,1}^2  \prod\limits_z \gF_{2\,1z}' \prod\limits_y \gF_{3\,1y}' 
+ \gF_{1\,2}^2 \prod\limits_z \gF_{2\,2z}' \prod\limits_y \gF_{3\,2y}' 
+ \gF_{1\,3}^2 \prod\limits_z \gF_{2\,3z}' \prod\limits_y \gF_{3\,3y}' 
\Big) \gG_1 
\\[1ex] & \phantom{\gk_1}+\,  
\Big( 
\gF_{1\,1}^2 \prod\limits_z \gF_{2\,1z}' \prod\limits_y \gF_{3\,1y}' 
+ \gF_{1\,2}^2 \prod\limits_z \gF_{2\,2z}' \prod\limits_y \gF_{3\,2y}' 
+ \gF_{1\,4}^2 \prod\limits_z \gF_{2\,4z}' \prod\limits_y \gF_{3\,4y}' 
\Big) \gG_1'
\\[2ex] & \,+ \, 
\Big( 
\gk_2\, \gF_{2\,1}^2  \prod\limits_x \gF_{1\,1x}' \prod\limits_z \gF_{3\,1z}' 
+ \gF_{2\,2}^2  \prod\limits_x \gF_{1\,2x}' \prod\limits_z \gF_{3\,2z}' 
+ \gF_{2\,3}^2  \prod\limits_x \gF_{1\,3x}' \prod\limits_z \gF_{3\,3z}' 
\Big) \gG_2 
\\[1ex] & \phantom{\gk_2}+ \, 
\Big( 
\gF_{2\,1}^2  \prod\limits_x \gF_{1\,1x}' \prod\limits_z \gF_{3\,1z}' 
+ \gF_{2\,2}^2  \prod\limits_x \gF_{1\,2x}' \prod\limits_z \gF_{3\,2z}' 
+ \gF_{2\,4}^2  \prod\limits_x \gF_{1\,4x}' \prod\limits_z \gF_{3\,4z}' 
\Big) \gG_2'
\\[2ex] & \,+ \,  
\Big( \gk_3\, \gF_{3\,1}^2  \prod\limits_x \gF_{1\,1x}' \prod\limits_y \gF_{2\,1y}' 
+ \gF_{3\,2}^2  \prod\limits_x \gF_{1\,2x}' \prod\limits_y \gF_{2\,2y}' 
+ \gF_{3\,3}^2  \prod\limits_x \gF_{1\,3x}' \prod\limits_y \gF_{2\,3y}' 
\Big) \gG_3 
\\[1ex] & \phantom{\gk_3}+ \, 
\Big( \gF_{3\,1}^2  \prod\limits_x \gF_{1\,1x}' \prod\limits_y \gF_{2\,1y}' 
+ \gF_{3\,2}^2  \prod\limits_x \gF_{1\,2x}' \prod\limits_y \gF_{2\,2y}' 
+ \gF_{3\,4}^2  \prod\limits_x \gF_{1\,4x}' \prod\limits_y \gF_{2\,4y}' 
\Big) \gG_3'
}
} 
in order to make it gauge invariant under all exceptional gaugings. The resulting D-- and F--term conditions are rather involved and not particularly illuminating. For this reason we refrain from giving them here and turn to the more transparant minimal full resolution model.

\subsection{Minimal full resolution GLSM}
\label{sc:MinFullResGLSM}

\begin{table} 
\[
\arry{|c||c|c|c||cc|cc|cc||c|c|c|}{
\hline 
\text{Superfield} & \gF_{1x} & \gF_{2y} & \gF_{3z} & \gG_1 & \gG_1' & \gG_2 & \gG_2' & \gG_3 & \gG_3' & \gF_{1}' & \gF_{2}' & \gF_{3}'  
\\ \hline 
\text{U(1) charge} & z_{1x} & z_{2y} & z_{3z} & \gg_1 & \gg_1' & \gg_2 & \gg_2' & \gg_3 & \gg_3' & x_{1} & x_{2} & x_{3}  
\\ \hline\hline 
R_1 & \sfrac 12 & 0 & 0 & -1 & -1 & 0 & 0 & 0 & 0 & 0 & 0 & 0 
\\ \hline 
R_2 & 0 & \sfrac 12 & 0 & 0 & 0 & -1 & -1 & 0 & 0 & 0 & 0 & 0 
\\ \hline 
R_3 & 0 & 0 & \sfrac 12 & 0 & 0 & 0 & 0 & -1 & -1 & 0 & 0 & 0 
\\ \hline\hline 
E_{1} & 0 & \sfrac 12 \gd_{y1} & \sfrac 12 \gd_{z1} & 0 & 0 & 0 & 0 & 0 & 0 & - 1 & 0 & 0 
\\ \hline 
E_{2} & \sfrac 12\gd_{x1} & 0 & \sfrac 12 \gd_{z1} & 0 & 0 & 0 & 0 & 0 & 0 & 0 & -1 & 0 
\\ \hline 
E_{3} & \sfrac 12 \gd_{x1} & \sfrac 12 \gd_{y1} & 0 & 0 & 0 & 0 & 0 & 0 & 0 & 0 & 0 & -1 
\\ \hline 
}
\]
\caption{ \label{tb:MinT6Z22glsm} 
A choice for a superfield charge table that determines the geometry of a minimal full resolution of $T^6/\Intr_2\times\Intr_2$.}
\end{table} 

The minimal full resolution GLSM has three ordinary and three exceptional gaugings. Contrary to the maximal full resolution GLSM, the charge assignments of minimal full resolution models are not unique as for each of the three exceptional gaugings there are $4\cdot 4 =16$ choices, which of the homogeneous coordinate superfields to be gauged.

Here only gaugings of the superfields $\gF_{1\, 1}$, $\gF_{2\, 1}$ and $\gF_{3\, 1}$ are considered\footnote{Other choices would be equally well justified, however we expect that the physical understanding does not depend much on this, even though the detailed description will.}, as can be seen in Table~\ref{tb:MinT6Z22glsm}. Consequently, the superpotential for the geometry reduces to 
\equ{ \label{eq:T6Z22minSuperpotential} 
P_\text{min\,res}  = 
 \sum\limits_{u=1}^3\gG_u \Big( 
\gk_u\, \gF_{u\,1}^2  \prod\limits_{r\neq u} \gF_{r}' 
+ \gF_{u\,2}^2 
+ \gF_{u\,3}^2
\Big)
\,+\,  
\sum\limits_{u=1}^3 \gG_u'\Big( 
\gF_{u\,1}^2 \prod\limits_{r\neq u} \gF_{2}'  \gF_{3}' 
+ \gF_{u\,2}^2 
+ \gF_{u\,4}^2
\Big)~. 
%
} 

The effective target space geometries are determined by six D-- and six F--term equations. The six resulting D--term conditions read 
\equ{ \label{eq:MinResDterms}
\arry{lcl}{
|z_{1\,1}|^2 + |z_{1\,2}|^2 + |z_{1\,3}|^2 + |z_{1\,4}|^2 = a_1~, 
& \quad\quad & 
|z_{2\,1}|^2 + |z_{3\,1}|^2 - 2\, |x_1|^2 = 2\, b_1~, 
\\[1ex] 
|z_{2\,1}|^2 + |z_{2\,2}|^2 + |z_{2\,3}|^2 + |z_{2\,4}|^2 = a_2~, 
& \quad\quad & 
|z_{1\,1}|^2 + |z_{3\,1}|^2 - 2\, |x_2|^2 = 2\, b_2~, 
\\[1ex] 
|z_{3\,1}|^2 + |z_{3\,2}|^2 + |z_{3\,3}|^2 + |z_{3\,4}|^2 = a_3~, 
& \quad\quad & 
|z_{1\,1}|^2 + |z_{2\,1}|^2 - 2\, |x_3|^2 = 2\, b_3
}
}
and the six F--term conditions 
\equ{ \label{eq:MinResFterms} 
\arry{lcl}{
\gk_1\, z_{1\,1}^2\, x_2x_3 + z_{1\,2}^2+z_{1\,3}^2 = 0~, 
& \quad\quad & 
z_{1\,1}^2\, x_2x_3 + z_{1\,2}^2+z_{1\,4}^2 = 0~, 
\\[1ex] 
\gk_2\, z_{2\,1}^2\, x_1x_3 + z_{2\,2}^2+z_{2\,3}^2 = 0~, 
& \quad\quad & 
z_{2\,1}^2\, x_1x_3 + z_{2\,2}^2+z_{2\,4}^2 = 0~, 
\\[1ex] 
\gk_3\, z_{3\,1}^2\, x_1x_2 + z_{3\,2}^2+z_{3\,3}^2 = 0~, 
& \quad\quad & 
z_{3\,1}^2\, x_1x_2 + z_{3\,2}^2+z_{3\,4}^2 = 0~. 
}
}
The properties of the resulting geometries depend crucially on the values of the K\"ahler parameters. As can be seen from the three D--term conditions on the left in \eqref{eq:MinResDterms} the parameters $a_1, a_2, a_3$ all need to be positive (since we have assumed that all $y^A=0$). The other K\"ahler parameters  $b_1, b_2, b_3$ may in principle have either sign.

\subsubsection*{Orbifold phase}

Consider the phase in which all three parameters  $b_1, b_2, b_3$ are negative while the parameters $a_1,a_2,a_3$ all positive. It follows that all three coordinates $x_1, x_2, x_3$ are necessarily non--zero so that their phases can be fixed to some preset values. This does not fix the gauge symmetries completely, as there are residual $\Intr_2$ actions left over: 
\equ{  \label{eq:ResidualZ2symmetries} 
\arry{l}{
\Intr_{2}:\quad z_{2\,1} \ra - z_{2\,1}~,\quad z_{3\,1} \ra - z_{3\,1}~, 
\\[1ex] 
\Intr_{2}:\quad z_{1\,1} \ra - z_{1\,1}~,\quad z_{3\,1} \ra - z_{3\,1}~, 
\\[1ex] 
\Intr_{2}:\quad z_{1\,1} \ra - z_{1\,1}~,\quad z_{2\,1} \ra - z_{2\,1}~.
}
} 
For concreteness, focus on the first of these three $\Intr_2$ actions. The fixed set of this action is given by $z_{2\,1} = z_{3\,1} = 0$. In the target space geometry this does not correspond to a single fixed set, but a collection of disjoint fixed sets. Indeed, inserting this in the second and third equations in \eqref{eq:MinResFterms} gives the equations: 
\equ{ 
z_{2\,2}^2+z_{2\,3}^2 = z_{2\,2}^2+z_{2\,4}^2 = 0~, 
\qquad 
z_{3\,2}^2+z_{3\,3}^2 = z_{3\,2}^2+z_{3\,4}^2 = 0~. 
}
Each of these equations are quadratic with two roots: 
\equ{ 
z_{2\,3} = \pm i\, z_{2\,2}~, 
\quad 
z_{2\,4} = \pm i\, z_{2\,2}~, 
\qquad 
z_{3\,3} = \pm i\, z_{3\,2}~, 
\quad 
z_{3\,4} = \pm i\, z_{3\,2}~, 
}
where all the signs are independent, hence there are $2^4 = 16$ solutions in total. Each of these fixed sets have the topology of a two--torus: The equations for the homogeneous coordinates $z_{1\,x}$ are those of the deformed two--torus used in Subsection~\ref{sc:TwoToriGLSM} since the absolute values of the coordinates $x_2$ and $x_3$ are determined by the second and third equation on the right hand side in \eqref{eq:MinResDterms}. This argumentation may be repeated for the second and third $\Intr_2$ actions in \eqref{eq:ResidualZ2symmetries}. Hence one has in total $3\cdot 16 = 48$ fixed two--tori; precisely the number of fixed two--tori to be expected in the $T^6/\Intr_2\times\Intr_2$ orbifold.

The coordinate patches suggested by the minimal full resolution model for the orbifold geometry can be extracted from the D-- and F--term equations~\eqref{eq:MinResDterms} and~\eqref{eq:MinResFterms}. Since all blowup parameters $b_1,b_2,b_3$ are negative, the three D--term equations on the right--hand--side of~\eqref{eq:MinResDterms} imply that $x_1,x_2,x_3\neq 0$. Each of the other three D--term equations imply that at least one coordinate in each is non--zero. But then the F--term equations~\eqref{eq:MinResFterms} imply that two other coordinates are non--zero. Hence, three out of four $z_{1x}$, $z_{2y}$ and $z_{3z}$ coordinates are non--zero. This leads to $4^3 = 64$ coordinate patches; the same number of coordinate patches as the $(T^2)^3$ torus GLSM would have.

\subsubsection*{Full resolution phases}

In the full resolution phases all parameters $b_1, b_2, b_3$ are positive but parametrically much smaller than the parameters $a_1, a_2, a_3$. (If this is not the case, the GLSM might develop more exotic phases, like critical-- and over--blowup phases~\cite{Blaszczyk:2011hs}.) In the full resolution phases it is useful to reshuffle the three D--term equations on the right hand side of \eqref{eq:MinResDterms} in the following fashion: 
\equ{  \label{eq:MixMinResDterms}
\arry{l}{
|z_{1\,1}|^2 + |x_1|^2 = b_2+b_3-b_1 + |x_2|^2 + |x_3|^2~. 
\\[1ex] 
|z_{2\,1}|^2 + |x_2|^2 = b_1+b_3-b_2 + |x_1|^2 + |x_3|^2~, 
\\[1ex] 
|z_{3\,1}|^2 + |x_3|^2 = b_1+b_2-b_3 + |x_1|^2 + |x_2|^2~. 
}
}
These equations contain important information as they decide which coordinate fields are necessarily non--zero. For example, if the sign of the combination $b_2+b_3-b_1$ is positive either $z_{1\,1}$ or $x_1$ is necessarily non--zero, while if this combination is negative either $x_2$ or $x_3$ is necessarily non--zero. 

The following divisors can be easily defined by setting one of the homogeneous coordinates to zero: the exceptional divisors $E_r := \{ x_r = 0 \}$ and the ordinary divisors $D_{u} := \{z_{u\, 1} = 0\}$. The exceptional divisors consists of $2^4 = 16$ disjoint components and the ordinary divisors of $2^2 = 4$ disjoint components. As was observed in~\cite{Blaszczyk:2011hs} the inherited torus divisors $R_u$ and $R_u'$ can be identified with the polynomials multiplying the chiral Fermi superfields $\gG_u$ and $\gG_u'$ in the superpotential~\eqref{eq:T6Z22minSuperpotential}.

\subsubsection*{S--triangulation full resolution phase} 

In the S--triangulation phase of the GLSM the three K\"ahler parameters are of similar size in the sense that each one is smaller than the sum of the other two, {\em e.g.}\ the following three inequalities 
\equ{ 
0 < b_1 < b_2+ b_3~, 
\quad 
0 < b_2 < b_1+ b_3~, 
\quad 
0 < b_3 < b_1+ b_2~,
}
hold simultaneously. In this phase the intersection $E_1E_2E_3$ exists because it is possible to satisfy all the D-- and F--term equations while setting $x_1=x_2=x_3=0$. In fact, in this case the F--term equations have $2^6 = 64$ solutions. This number comes as no surprise, since the $T^6/\Intr_2\times\Intr_2$ has 64 $\Intr_2\times\Intr_2$ fixed points. When all resolved using the S--triangulation, one 64 times the intersection of these three exceptional divisors. Note that the first equation in \eqref{eq:MixMinResDterms} implies that not both $z_{1\,1}$ and $x_1$ can be zero at the same time, hence, in particular, the curve $D_1 E_1$ does not exist. All this is in accordance with the topological properties of the S--triangulation of the resolved $T^6/\Intr_2\times \Intr_2$ orbifold.

\subsubsection*{E$_1$--triangulation full resolution phase} 

In the E$_1$--triangulation phase of the GLSM the K\"ahler parameter $b_1$ is much larger than the sum of the other two: 
\equ{
0 < b_2 < b_1+ b_3~, 
\quad 
0 < b_3 < b_1+ b_2~, 
\quad
 b_2+ b_3 < b_1~. 
}
Then \eqref{eq:MixMinResDterms} implies that not both $x_2$ and $x_3$ can be zero at the same time, hence, in particular, the curve $E_2E_3$ and the intersection $E_1E_2E_3$ do not exist in this phase. Contrary, in this phase the curve $D_1E_1$ does exist. All this is, again, in accordance with the topological properties of the E$_1$--triangulation of the resolved $T^6/\Intr_2\times\Intr_2$ orbifold. The transition from the S-- to the E$_1$--triangulation phase thus provides the GLSM description of the flop transition. Note that in the GLSM there is nothing singular at the transition $b_1=b_2+b_3$ even though the target space geometry is singular there.

The other two full resolutions phases, the E$_2$-- and $E_3$--triangulations may be defined in an analogous fashion.

\subsubsection*{Full resolution coordinate patches}

To understand the coordinate patches in the full resolution phases, first observe that the three D--term equations on the right--hand--side of~\eqref{eq:MinResDterms} lead to the same options for non--vanishing coordinates $z_{1\,1},z_{2\,1},z_{3\,1},x_1,x_2,x_3$ as obtained in the non--compact case summarised in Table~\ref{tb:C3Z22Phases}. Hence, in the $S$--, $E_1$--, $E_2$-- or $E_3$--triangulation the following coordinate combinations 
\sequ{eq:NonVanishingCoordsExcpt}{
S:~~z_{1\,1}z_{2\,1}z_{3\,1} \neq 0~, 
~~ 
z_{1\,1}z_{2\,1}x_3 \neq 0~, 
~~ 
z_{1\,1}z_{3\,1}x_2 \neq 0
~~\text{or}~~ 
z_{2\,1}z_{3\,1}x_1 \neq 0~, 
\\[1ex] 
E_1:~~z_{2\,1}z_{3\,1}x_3 \neq 0~, 
~~ 
z_{1\,1}z_{2\,1}x_3 \neq 0~, 
~~ 
z_{2\,1}z_{3\,1}x_2 \neq 0
~~\text{or}~~ 
z_{1\,1}z_{3\,1}x_2 \neq 0~, 
\\[1ex] 
E_2:~~z_{1\,1}z_{3\,1}x_3 \neq 0~, 
~~ 
z_{1\,1}z_{2\,1}x_3 \neq 0~, 
~~ 
z_{2\,1}z_{3\,1}x_1 \neq 0
~~\text{or}~~ 
z_{1\,1}z_{3\,1}x_1 \neq 0~, 
\\[1ex] 
E_3:~~z_{2\,1}z_{3\,1}x_1 \neq 0~, 
~~ 
z_{1\,1}z_{2\,1}x_1 \neq 0~, 
~~ 
z_{1\,1}z_{2\,1}x_2 \neq 0
~~\text{or}~~ 
z_{1\,1}z_{3\,1}x_2 \neq 0~. 
}
are non--zero, respectively. Next, observe that the first D--term equation on the left--hand--side of~\eqref{eq:MinResDterms} implies that at least $z_{1\,x}$ is non--zero. If this happens to be $z_{1\,1}$ then the latter two D--term equations on the right--hand--side of~\eqref{eq:MinResDterms} imply that $x_2$ and $x_3$ are also non--zero because $a_1$ is parametrically larger than the parameters $b_1,b_2,b_3$ so that cancellations are never possible. But the the two top F--term equations~\eqref{eq:MinResFterms} imply that two other $z_{1x}$, $x\neq1$ are non--zero. There are three options for this to happen. Finally, it is possible that all three $z_{1\,x}$, $x\neq 1$ are non--zero. In total this gives four non--vanishing coordinate combinations for the first lines of the D-- and F--term equations. A similar analysis can be performed for the second and third lines of these equations, leading to the following combinations of non--vanishing coordinates 
\sequ{eq:NonVanishingCoords}{
z_{1\,2}z_{1\,3}z_{1\,4}\neq 0~, 
~~ 
z_{1\,1}z_{1\,3}z_{1\,4}x_2x_3\neq 0~, 
~~ 
z_{1\,1}z_{1\,2}z_{1\,4}x_2x_3\neq 0
~~\text{or}~~ 
z_{1\,1}z_{1\,2}z_{1\,3}x_2x_3\neq 0~, 
\\[1ex] 
z_{2\,2}z_{2\,3}z_{2\,4}\neq 0~, 
~~ 
z_{2\,1}z_{2\,3}z_{2\,4}x_1x_3\neq 0~, 
~~ 
z_{2\,1}z_{2\,2}z_{2\,4}x_1x_3\neq 0
~~\text{or}~~ 
z_{2\,1}z_{2\,2}z_{2\,3}x_1x_3\neq 0~, 
\\[1ex] 
z_{3\,2}z_{3\,3}z_{3\,4}\neq 0~, 
~~ 
z_{3\,1}z_{3\,3}z_{3\,4}x_1x_2\neq 0~, 
~~ 
z_{3\,1}z_{3\,2}z_{3\,4}x_1x_2\neq 0
~~\text{or}~~
z_{3\,1}z_{3\,2}z_{3\,3}x_1x_2\neq 0~.
}
Coordinate patches can now be composed by taking one out of four equations on each line of~\eqref{eq:NonVanishingCoords} combined with one out of the four equations from the line in~\eqref{eq:NonVanishingCoordsExcpt} corresponding to the chosen triangulation. Not all combinations are valid however, in total there should be 12 non--vanishing coordinates out of the 15 original ones, so that the coordinate patch has complex dimension three. 

The results of this analysis are summarised in Table~\ref{tb:CoordinatePatchesCompact}. The GLSM description leads to 76 coordinate patches for each of the full resolution phases. There are 72 universal coordinate patches which exist independently of which triangulation is chosen: 
for each triangulation choice in~\eqref{eq:NonVanishingCoordsExcpt} there is at least one combination of non--vanishing fields which is contained in the non--vanishing set coordinates of that patch to the extent that precisely 12 coordinates are non--zero.  
54 of those patches do not involve any of the exceptional coordinates and therefore coincide with the coordinate patches of the orbifold discussed above. These coordinate patches are indicated above the line that splits the universal patches in Table~\ref{tb:CoordinatePatchesCompact}. In addition, to the 72 universal coordinate patches there are four patches that depend on the triangulation. The GLSM therefore dictates a gluing procedure in which ten of the coordinate patches of the orbifold are replaced by 22 patches for the full resolutions.

\begin{table} 
\begin{center} 
\begin{tabular}{|c||c|c|c|c|}
\hline 
Phase & $\#$ &Non--zero fields  & Patches & Conditions
\\\hline\hline
Universal & $54$ & $z_{1\,x'\neq x}\,z_{2\,y'\neq y}\,z_{z'\neq z}\,x_1\,x_2\,x_3 \neq 0$ & $\{z_{1\,x},z_{2\,y},z_{3\,z}\}$ & $x,y,z\neq 1$ 
\\ 
&  &  $z_{1\,x'\neq 1}\,z_{2\,y'\neq y}\,z_{z'\neq z}\,x_1\,x_2\,x_3 \neq 0$ & $\{z_{1\,1},z_{2\,y},z_{3\,z}\}$ & $y,z\neq 1$ 
\\
&  & $z_{1\,x'\neq x}\,z_{2\,y'\neq 1}\,z_{z'\neq z}\,x_1\,x_2\,x_3 \neq 0$ & $\{z_{1\,x},z_{2\,1},z_{3\,z}\}$ & $x,z\neq 1$ 
\\
&  & $z_{1\,x'\neq x}\,z_{2\,y'\neq y}\,z_{z'\neq 1}\,x_1\,x_2\,x_3 \neq 0$ & $\{z_{1\,x},z_{2\,y},z_{3\,1}\}$ & $x,y\neq 1$ 
\\ \cline{2-5} 
& $18$ &  $z_{u'\neq u\,1}z_{1\,x'\neq 1}\,z_{2\,y'\neq 1}\,z_{z'\neq z}\,x_1\,x_2 \neq 0$ & $\{z_{u\,1},z_{3\,z},x_3\}$ & $u',u=1,2; z\neq 1$ 
\\ 
&  &  $z_{u'\neq u\,1}z_{1\,x'\neq 1}\,z_{2\,y'\neq y}\,z_{z'\neq 1}\,x_1\,x_3 \neq 0$ & $\{z_{u\,1},z_{2\,y},x_2\}$ & $u',u\neq1,3; y\neq 1$ 
\\
&  &  $z_{u'\neq u\,1}z_{1\,x'\neq x}\,z_{2\,y'\neq 1}\,z_{z'\neq 1}\,x_2\,x_3 \neq 0$ & $\{z_{u\,1},z_{1\,x},x_{1}\}$ & $u',u=2,3; x\neq 1$ 
\\\hline\hline 
S--triang. & $4$ & $z_{u\,x}\neq 0$ & $\{x_1,x_2,x_3\}$ & $u=1,2,3; x=1,..,4$ 
\\
 &  & $z_{u\,x \neq 3\,1}x_3\neq 0$ & $\{z_{3\,1},x_1,x_2\}$ & $u=1,2,3; x=1,..,4$  
\\\
 &  & $z_{u\,x \neq 2\,1}x_2\neq 0$ & $\{z_{2\,1},x_1,x_3\}$ & $u=1,2,3; x=1,..,4$  
\\
 &  & $z_{u\,x \neq 1\,1}x_1\neq 0$ & $\{z_{1\,1},x_2,x_3\}$ & $u=1,2,3; x=1,..,4$  
\\\hline\hline 
E$_1$--triang.  & $4$ & $z_{u\,x\neq 1\,1}x_3\neq 0$ & $\{z_{1\,1},x_1,x_2\}$ & $u=1,2,3; x=1,..,4$ 
\\
&  & $z_{u\,x \neq 3\,1}x_3\neq 0$ & $\{z_{3\,1},x_1,x_2\}$ & $u=1,2,3; x=1,..,4$  
\\ 
 &  & $z_{u\,x \neq 1\,1}x_2\neq 0$ & $\{z_{1\,1},x_1,x_3\}$ & $u=1,2,3; x=1,..,4$  
\\
 &  & $z_{u\,x \neq 2\,1}x_2\neq 0$ & $\{z_{2\,1},x_1,x_3\}$ & $u=1,2,3; x=1,..,4$  
\\\hline\hline 
E$_2$--triang.  & $4$ & $z_{u\,x\neq 2\,1}x_3\neq 0$ & $\{z_{2\,1},x_1,x_2\}$ & $u=1,2,3; x=1,..,4$ 
\\
&  & $z_{u\,x \neq 3\,1}x_3\neq 0$ & $\{z_{3\,1},x_1,x_2\}$ & $u=1,2,3; x=1,..,4$  
\\ 
 &  & $z_{u\,x \neq 1\,1}x_1\neq 0$ & $\{z_{1\,1},x_2,x_3\}$ & $u=1,2,3; x=1,..,4$  
\\
 &  & $z_{u\,x \neq 2\,1}x_1\neq 0$ & $\{z_{2\,1},x_2,x_3\}$ & $u=1,2,3; x=1,..,4$  
\\\hline\hline 
E$_3$--triang.  & $4$ & $z_{u\,x\neq 1\,1}x_1\neq 0$ & $\{z_{1\,1},x_2,x_3\}$ & $u=1,2,3; x=1,..,4$ 
\\
&  & $z_{u\,x \neq 3\,1}x_1\neq 0$ & $\{z_{3\,1},x_2,x_3\}$ & $u=1,2,3; x=1,..,4$  
\\ 
 &  & $z_{u\,x \neq 3\,1}x_2\neq 0$ & $\{z_{3\,1},x_1,x_3\}$ & $u=1,2,3; x=1,..,4$  
\\
 &  & $z_{u\,x \neq 2\,1}x_2\neq 0$ & $\{z_{2\,1},x_1,x_3\}$ & $u=1,2,3; x=1,..,4$  
\\\hline 
\end{tabular}
\end{center} 
\caption{ \label{tb:CoordinatePatchesCompact} 
The 76 coordinate patches of the full resolution phases of the minimal full resolution GLSM. There are 72 universal coordinate patches which are the same for each of the full resolution phases. In addition, there are four coordinate patches which are specific for the triangulation chosen.}
\end{table}

\subsection{Gauge background on the minimal full resolution of the non--torsional orbifold}
\label{sc:GaugeBackgroundMinFullRes}

The gauge charges of the Fermi and chiral multiplets that define a simple gauge bundle on the minimal full resolution model is given in Table~\ref{tb:MinT6Z22bundle}. This gauge bundle is quite closely related to the standard embedding on the two--tori. The exceptional $E$--gauge charges are identical to those indicated in \eqref{eq:AnomalyCancellationC3Z22} of the non--compact resolution model. In order to avoid any of the four types of anomalies mentioned in Subsection~\ref{sc:AnomalyConsistencyConditions}, additional chiral multiplets $\gPs_u, \gPs_u'$ are introduced with identical charges as $\gG_u, \gG_u'$ and the sum of charges of all chiral superfields and all chiral Fermi superfields vanish separately.

In total there are $3\cdot 4+ 3 = 15$ Fermi multiplets involved in the gauge bundle subject to $3\cdot 2 = 6$ constraints enforced by the chiral multiplets $\gPs_u, \gPs_u'$. This leave nine Fermi multiplets part of the gauge background which cannot be fitted into a single $E_8$ factor. Hence a number of fermionic gaugings are needed. If all six gaugings in the minimal resolution model are accompanied by fermionic gaugings, (a deformation of) the standard embedding is obtained. To make contact with the non--torsion line bundle model that was discussed in Section~\ref{sc:PairsTorsionGLSMs}, only the inherited $R_u$--gaugings are accompanied with fermionic gaugings with parameters $\gX_u$, while the exceptional $E_r$--gaugings are not. With this choice of Fermionic gaugings $9-3=6$ gauge bundle directions are left over, exactly matching the number in the non--compact resolution of the non--torsion orbifold model. 

In target space this gauge background does not correspond to the standard embedding as there are no fermionic gaugings associated to the exceptional $E_r$--gaugings. Neither can this background be interpreted as line bundles only because of the presence of the chiral multiplets $\gPs_u, \gPs_u'$ that enforce constraints on the bundle degrees of freedom as well as the fermionic gaugings $\gX_u$.

\begin{table} 
\[
\arry{|c||c|c|c||cc|cc|cc||c|c|c||c|}{
\hline 
\text{Superfield} & \gL_{1x} & \gL_{2y} & \gL_{3z} & \gPs_1 & \gPs_1' & \gPs_2 & \gPs_2' & \gPs_3 & \gPs_3' & \gL_{1}' & \gL_{2}' & \gL_{3}'  & \gL_n
\\ \hline 
\text{U(1) charge} & \gl_{1x} & \gl_{2y} & \gl_{3z} & \gps_1 & \gps_1' & \gps_2 & \gps_2' & \gps_3 & \gps_3' & \gl_{1}' & \gl_{2}' & \gl_{3}'  & \gl_n
\\ \hline\hline 
R_1 & \sfrac 12 & 0 & 0 & -1 & -1 & 0 & 0 & 0 & 0 & 0 & 0 & 0 & 0
\\ \hline 
R_2 & 0 & \sfrac 12 & 0 & 0 & 0 & -1 & -1 & 0 & 0 & 0 & 0 & 0 & 0 
\\ \hline 
R_3 & 0 & 0 & \sfrac 12 & 0 & 0 & 0 & 0 & -1 & -1 & 0 & 0 & 0 & 0 
\\ \hline\hline 
E_{1} & 0 & \sfrac 12 \gd_{y1} & \sfrac 12 \gd_{z1} & 0 & 0 & 0 & 0 & 0 & 0 & - 1 & 0 & 0 & 0
\\ \hline 
E_{2} & \sfrac 12\gd_{x1} & 0 & \sfrac 12 \gd_{z1} & 0 & 0 & 0 & 0 & 0 & 0 & 0 & -1 & 0 & 0
\\ \hline 
E_{3} & \sfrac 12 \gd_{x1} & \sfrac 12 \gd_{y1} & 0 & 0 & 0 & 0 & 0 & 0 & 0 & 0 & 0 & -1 & 0
\\ \hline 
}
\]
\caption{ \label{tb:MinT6Z22bundle} 
A choice for a charge table of the superfields that determine a gauge bundle on the minimal full resolution of $T^6/\Intr_2\times\Intr_2$. The Fermi multiplets $\gL_n$, $n=1,\ldots, 18$, are spectators and generate the broken gauge group.}
\end{table}

Given the charges of Table~\ref{tb:MinT6Z22bundle} the following superpotential can be written down: 
%
\equ{ \label{eq:SuperpotentialT6Z22bundleExplicit} 
\arry{rl}{
P_\text{min\,res\,bundle} & = 
\sum\limits_{u=1}^3 \gPs_u \Big( 
2\gk_u\, \gF_{u\,1}  \prod\limits_{r\neq u} \gF_{r}'\, \gL_{u\,1}
+ \gk_u\, \gF_{u\,1}^2  \prod\limits_{r\neq s \neq u}  \gF_{r}' \gL_{s}' 
+ 2\, \gF_{u\,2} \gL_{u\,2}  
+ 2 \, \gF_{u\,3} \gL_{u\,3}
\Big)
\\[2ex] &\,+\,  
\sum\limits_{u=1}^3 \gPs_u'\Big( 
2\, \gF_{u\,1}  \prod\limits_{r\neq u} \gF_{r}'\, \gL_{u\,1}
+  \gF_{u\,1}^2   \prod\limits_{r\neq s \neq u}  \gF_{r}' \gL_{s}' 
+ 2\, \gF_{u\,2} \gL_{u\,2}  
+ 2\, \gF_{u\,4} \gL_{u\,4}
\Big)~.
}
} 
This specific form of a general expression for this superpotential is inspired by the standard embedding following~\eqref{sc:StandardEmbeddingSuperpotential}.

In the model under investigation only the $R_u$--gaugings are associated to fermionic gauge transformations, hence the only non--zero fermionic gauge transformations are: 
\equ{ \label{eq:FermiGaugeBundle} 
\gd \gL_{u\,x} = \sfrac 12\, \gF_{u\,x} \,\gX_u~, 
\qquad  
\gd \gG_u = - \gPs_u\, \gX_u~, 
\qquad 
\gd \gG_u' = - \gPs_u'\, \gX_u~. 
%
}
The specific form, given here, is obtained by requiring that the fermionic gauges are on the (2,2)--locus. In this case is follows automatically that \eqref{eq:T6Z22minSuperpotential} and \eqref{eq:SuperpotentialT6Z22bundleExplicit}  combined are inert under these fermionic transformations.

 This construction leads to a regular bundle as for each of the three fermionic gaugings in~\eqref{eq:FermiGaugeBundle} not all coefficients vanish simultaneously. The same goes for the six constraints coming from~\eqref{eq:SuperpotentialT6Z22bundleExplicit}. It is straightforward to check this for all coordinate patches given in Table~\ref{tb:CoordinatePatchesCompact} for all four fully resolved phases of this GLSM. This should not come as a surprise as the fermionic gaugings~\eqref{eq:FermiGaugeBundle} and the bundle superpotential~\eqref{eq:SuperpotentialT6Z22bundleExplicit} are precisely those that are dictated by the (2,2) locus, see Subsection~\ref{sc:22Locus}.

\subsection{Gauge background on the minimal full resolution of the torsional orbifold}
\label{sc:GaugeBackgroundMinFullResTorsion}

In section~\ref{sc:OrbifoldSpectraWithoutTorsion} it was explained that the twisted states that survive the orbifold projections are precisely opposite when torsion is switched on to when it is absent. Since the shifted momenta of the twisted states without oscillators dictated the exceptional $E_1, E_2, E_3$--charges in the GLSM of the Fermi multiplet $\gL$. Hence the charge Table~\ref{tb:MinT6Z22glsm}, which determines the geometry, remains unchanged when torsion is switched on, but the charge table for the vector bundle is modified to Table~\ref{tb:MinT6Z22bundleTorsion}: the $R_i$--charges remain the same while the $E_r$--charges are all sign--flipped as compared to those in Table~\ref{tb:MinT6Z22bundle}.

\begin{table} 
\[
\arry{|c||c|c|c||cc|cc|cc||c|c|c||c|}{
\hline 
\text{Superfield} & \gL_{1x}^{\!\!\times} & \gL_{2y}^{\!\!\times} & \gL_{3z}^{\!\!\times} & \gPs_1 & \gPs_1' & \gPs_2 & \gPs_2' & \gPs_3 & \gPs_3' & \gL_{1}^{\!\!\times\,\prime} & \gL_{2}^{\!\!\times\,\prime} & \gL_{3}^{\!\!\times\,\prime}  & \gL_{n}
\\ \hline 
\text{U(1) charge} & \gl_{1x} & \gl_{2y} & \gl_{3z} & \gps_1 & \gps_1' & \gps_2 & \gps_2' & \gps_3 & \gps_3' & \gl_{1}' & \gl_{2}' & \gl_{3}'  & \gl_n 
\\ \hline\hline 
R_1 & \sfrac 12 & 0 & 0 & -1 & -1 & 0 & 0 & 0 & 0 & 0 & 0 & 0 & 0
\\ \hline 
R_2 & 0 & \sfrac 12 & 0 & 0 & 0 & -1 & -1 & 0 & 0 & 0 & 0 & 0 & 0 
\\ \hline 
R_3 & 0 & 0 & \sfrac 12 & 0 & 0 & 0 & 0 & -1 & -1 & 0 & 0 & 0 & 0
\\ \hline\hline 
E_{1} & 0 & -\sfrac 12 \gd_{y1} & -\sfrac 12 \gd_{z1} & 0 & 0 & 0 & 0 & 0 & 0 & +1 & 0 & 0 & 0 
\\ \hline 
E_{2} & -\sfrac 12\gd_{x1} & 0 & -\sfrac 12 \gd_{z1} & 0 & 0 & 0 & 0 & 0 & 0 & 0 & +1 & 0 & 0
\\ \hline 
E_{3} & -\sfrac 12 \gd_{x1} & -\sfrac 12 \gd_{y1} & 0 & 0 & 0 & 0 & 0 & 0 & 0 & 0 & 0 & +1 & 0 
\\ \hline 
}
\]
\caption{ \label{tb:MinT6Z22bundleTorsion} 
A choice for a charge table of the superfields that determine a gauge bundle on the minimal full resolution of $T^6/\Intr_2\times\Intr_2$ with torsion.}
\end{table}

The flipping of the $E_r$--gauge charges has various consequences. First of all, the fermionic gauge transformations~\eqref{eq:FermiGaugeBundle} are not gauge covariant any more. This is easily alleviated by inserting appropriate factors of $\gF_r'$ in the first column of fermionic gauge transformations of $\gL_{u1}$: 
\equ{ \label{eq:FermiGaugeBundleTorsion} 
\gd \gL_{u\,1}^{\!\!\times} = \sfrac 12\, \prod_{r\neq u}\gF_r'\, \gF_{u\,1} \,\gX_u~, 
\qquad  
\gd \gL_{u\,x}^{\!\!\times} = \sfrac 12\, \gF_{u\,x} \,\gX_u~, 
\qquad 
\gd \gG_u = - \gPs_u\, \gX_u~, 
\qquad 
\gd \gG_u' = - \gPs_u'\, \gX_u~, 
%
}
for $x \neq 1$. Secondly, the bundle superpotential~\eqref{eq:SuperpotentialT6Z22bundleExplicit}  has to be modified to 
\equ{ \label{eq:SuperpotentialT6Z22bundleExplicitTorsion} 
\arry{rl}{
P_\text{min\,res\,bundle} & = 
 \sum\limits_{u=1}^3 \gPs_u \Big( 
2\gk_u\, \gF_{u\,1}  \, \gL_{u\,1}^{\!\!\times}
+ \gk_u\, \gF_{u\,1}^2  \prod\limits_{r\neq s\neq u}  \gF_{r}' \gF_s^{\prime 2} \gL_{s}^{\!\!\times\,\prime}
+ 2\, \gF_{u\,2} \gL_{u\,2}^{\!\!\times}  
+ 2 \, \gF_{u\,3} \gL_{u\,3}^{\!\!\times}
\Big)
\\[2ex] &\,+\,  
 \sum\limits_{u=1}^3 \gPs_u'\Big( 
2\, \gF_{u\,1}  \, \gL_{u\,1}^{\!\!\times}
+  \gF_{u\,1}^2  \prod\limits_{r\neq s\neq u}  \gF_{r}' \gF_s^{\prime 2} \gL_{s}^{\!\!\times\,\prime}
+ 2\, \gF_{u\,2} \gL_{u\,2}^{\!\!\times}  
+ 2\, \gF_{u\,4} \gL_{u\,4}^{\!\!\times}
\Big)
}
} 
by making the following replacements 
\equ{ \label{eq:ReplacementsTorsion} 
\gL_{u1} \ra \gF_r^{\prime -1} \gF_{s}^{\prime -1} \gL_{u1}^{\!\!\times}~, 
\quad 
\gL_{ux} \ra \gL_{ux}^{\!\!\times}~, 
\quad
\gL_r^\prime \ra \gF_r^{\prime 2} \gL_r^{\!\!\times\,\prime}~, 
}
with $x \neq 1$ and $r\neq s\neq u$, to ensure that it is gauge invariant again. With these modifications of the fermionic gauge transformations and the bundle superpotential, it is not difficult to see that the full superpotential including the part for the geometry \eqref{eq:T6Z22maxSuperpotential}  is invariant under the fermionic gauge transformations.

The replacements~\eqref{eq:ReplacementsTorsion} are the same as the field redefinitions~\eqref{eq:SuperfieldRedefinition} in the non--compact case with the chiral superfields $\gO_r$ given by the ones in the orbifold case ($O$) of Table~\ref{tb:OmegaRepresentationsC3Z22Phases}. It should be stressed that in the present case the replacements~\eqref{eq:ReplacementsTorsion} in the bundle superpotential apply to the GLSM theory as a whole globally, not just to a particular (coordinate patch within a) phase of the theory. Moreover, it is unique in the sense that other factors, that would have the same charges (like the other combinations in Table~\ref{tb:OmegaRepresentationsC3Z22Phases}), would always involve some powers of $\gPs_u$ or $\gPs_u'$, but that is forbidden because they are only allowed to appear linearly in the superpotential because of the R--symmetry, as was emphasised below~\eqref{eq:SPfermigaugeInv}.

\subsubsection*{Mixed anomalies and worldsheet Green--Schwarz mechanism}

The flipped $E_r$--gauge charges in Table~\ref{tb:MinT6Z22bundleTorsion} is irrelevant for most anomalies which still vanish identically as can be verified using~\eqref{eq:GaugeAnomalies} through~\eqref{eq:rightWeylAnomalies}. Only mixed $R_u E_{r\neq u}$--anomalies are now non--zero: 
\equ{
\cA_{ur} = \cA_{ru} = \sfrac 12\cdot \sfrac 12 - \sfrac12\cdot (-\sfrac 12) = \sfrac 12~, 
} 
$u \neq r$. Hence, contrary to the GLSMs associated to the non--compact orbifold models, the GLSMs associated to the compact orbifold models without or with torsion are genuinely physically distinct.

These mixed anomalies need to be cancelled by field dependent FI--terms of the form
\equ{ \label{eq:FIanom} 
W_\text{FI\,anom} = \frac 1{4\pi} \sum_{u,r} \frac{c_{ru}}2\, \log (N^r)\, F^u + 
\frac 1{4\pi} \sum_{u,r}\frac{1-c_{ru}}2 \,\log (N^u)\, F^r~, 
}
where the composite $N^r$ and $N^u$ have negative unit charge under the $R_u$-- and $E_r$--gaugings, respectively, and all other gauge charges zero. The arbitrary coefficients $c_{ru}$ arise as it is possible by counter terms to shift two dimensional mixed anomalies around. The choice $c_{ru} = 1/2$ would treat all mixed anomalies symmetrically. (See {\em e.g.}\ ref.~\cite{Blaszczyk:2011ib} for a more extensive discussion.) The composite chiral superfields $N^u$ and $N^r$ can be realised as rational functions of (fractional) powers of the chiral superfields. They may be expressed as 
%
\equ{
N^r = \gF_r'
~,
\quad 
N^u = 
\sum_{x,y\neq 1} n_{uxy}\, \gF_{ux}^{-1} \gF_{uy}^{-1} + n_{u11} \gF_{u1}^{-2} \prod_{r\neq u}\gF_r^{\prime -1} + 
\sum_{x\neq 1} n_{u1x}\, \gF_{u1}^{-1} \gF_{ux}^{-1} \prod_{r\neq u} \gF_r^{\prime -\sfrac 12}
~, 
}
%
with some generically non--zero parameters $n_{uxy}$, $n_{u11}$ and $n_{u1x}$. Since the chiral superfields $\gPs_u$ and $\gPs_u'$ cannot appear here as they would break $\text{R}$--symmetry, the possible forms in these expressions are restricted.

The superfield dependent FI--terms~\eqref{eq:FIanom} are defined on the level of the definition of the model and are singular independently of how the coefficients $c_{ru}$ and  $n_{uxy}$, $n_{u11}$ and $n_{u1x}$ are chosen, hence they signify the presence of NS5--branes~\cite{Blaszczyk:2011ib,Quigley:2011pv}. The interpretation of the coefficients $c_{ru}$ is not so clear. However, if they are all set to zero: $c_{ru} = 1$, then the expressions of $N^u$ become irrelevant. The NS5--branes are then located on the resolved exceptional cycles $E_r$ and they would disappear inside the orbifold singularities in the blow down limit. Maybe other values of $c_{ru}$ could be interpreted that the NS5--branes are moved around the resolved orbifold geometry and for $c_{ru}=0$ they are pushed fully off the resolved singularities onto the two--torus cycles. This seems to signify that the NS5--branes can move around on the resolved geometry without losing their influx effects on the worldsheet. This interpretation may be more transparent in another parameterisation 
\equ{
W_\text{FI anom} = 
\frac 1{8\gp} \sum_{u,r} c_{ru} \log \gF_r' F^u - 
\frac 1{4\gp} \sum_{u,r} \!\left[ \sum_{x\neq 1} c_{uxr} \log \gF_{u\,x}  + 
c_{u1r} \Big( \log \gF_{u\,1} \!-\! \sum_{r'\neq u}\sfrac 12\log\gF_{r'}'\Big)\!
\right]\!\! F^r
}
of~\eqref{eq:FIanom}, since the coefficients determining the position of the NS5--branes are subject to the constraint 
\(
c_{ru} + \sum\limits_{x} c_{uxr}  = 1\,.
\)

\subsubsection*{Comparing the pair of torsion related GLSMs}

Just like in the non--compact case, it is instructive to compare the resoluton GLSMs of the orbifold theories without and with torsion with each other by working in the same superfield basis. By interpreting~\eqref{eq:ReplacementsTorsion} as a superfield redefinition~\eqref{eq:AnomSuperfieldRedefinition}, but now both $R_u$-- and $E_u$--transformations are involved, the anomaly matrix $\cA$ extends to 
\equ{ 
\cA = \pmtrx{
~\sfrac 14~ & 0 & 0 & 0 & ~\sfrac 14~ & ~\sfrac 14~ 
\\ 
0 & ~\sfrac 14~ & 0 & ~\sfrac 14~ & 0 & ~\sfrac 14~
\\
0 & 0 & ~\sfrac 14~ & ~\sfrac 14~ & ~\sfrac 14~ & 0
\\
0 & ~\sfrac 14~ & ~\sfrac 14~  & ~\sfrac 32~ & ~\sfrac 14~ & ~\sfrac 14~ 
\\
~\sfrac 14~ & 0 & ~\sfrac 14~ &  \sfrac 14  &\sfrac 32 & \sfrac 14 
 \\  
~\sfrac 14~ & ~\sfrac 14~ & 0 &  \sfrac 14 & \sfrac 14 &  \sfrac 32 
}~.  
}
Notice that the lower $3\times 3$--block is identical to~\eqref{eq:AnomMatrixSuperfieldRedefinition}. Since in the replacements~\eqref{eq:ReplacementsTorsion} only the superfields $\gF_r'$ feature, the superfield redefintion anomalies reads 
\equ{ 
W_\text{field\, redef\, anom} = 
-\frac 1{2\gp} \, 
\Big\{ 
\frac 14 \sum_{u\neq r} \log\gF_r' \, F^u 
+ \frac 32 \sum_r  \log \gF_r' \, F^r 
+  \frac 14 \sum_{r'\neq r} \log\gF_r' \, F^{r'} 
\Big\}~. 
} 
The first contributions coincides with the general expression~\eqref{eq:FIanom} provided that $c_{ur} = 0$ and $N^r = \gF_r'$, hence they cancel each other exactly. The latter two contributions were also obtained in the non--compact situation~\eqref{eq:AnomSuperfieldRedefinition}. Hence, the analysis performed in Subsection~\eqref{sc:PairsTorsionGLSMs} can be repeated here as well. In particular in the orbifold phase, that analysis recovers the discrete torsion phases.

\setcounter{equation}{0}
\section{Conclusions} 
\label{sc:Conclusions}

Discrete torsion within the $\Intr_2\times\Intr_2$ orbifolds correspond to particular additional phases between the sum of partition functions of different sectors corresponding to different boundary conditions on the worldsheet torus. Smooth geometries are typically described by NLSMs which cannot be exactly quantised and the path integral cannot be represented as a sum over similar sectors as the orbifold theory. It is therefore unclear how to include effects of discrete torsion for smooth geometries. The main aim of this paper was to understand where discrete torsion goes when orbifolds have been resolved to fully smooth geometries. This question was addressed both for resolutions of the non--compact orbifold $\Cplx^3/\Intr_2\times\Intr_2$ as well as the compact $T^6/\Intr_2\times\Intr_2$ orbifold with Hodge numbers $(51,3)$ to understand both local and global aspects.

GLSMs were chosen as the framework for this investigation, as they can both make contact with the orbifolds as well as with fully resolved smooth geometries within the same description. From an effective field theory point of view orbifold resolutions correspond to giving VEVs to twisted states defining the blowup modes. Unless very particular blowup modes are selected, this leads to $(0,2)$ compactifications in which the gauge backgrounds are not dictated by the standard embedding. Therefore, in this work $(0,2)$ GLSMs were used for the interpolation between singular orbifolds and smooth compactifications.

The non--compact resolution GLSM of the $\Cplx^3/\Intr_2\times\Intr_2$ geometry had already given in the literature, the same goes  for the resulting line bundle backgrounds obtained by using non--oscillator blowup modes on the three $\Cplx^2/\Intr_2$ singularities. The GLSM gauge charges of the chiral Fermi multiplets under the resulting three exceptional gauge symmetries are given as the shifted left--moving momenta of these blowup modes. The effect of discrete torsion on the orbifold is that the twisted states with the opposite left--moving shifted momenta survive the orbifold projections. Consequently, the chiral Fermi multiplets in resolution GLSM for the torsional orbifold has the opposite worldsheet gauge charges as the non--torsional case. The GLSM associated to the torsional orbifold is equally well defined as the non--torsional model in the sense that all (gauge) anomalies vanish. In many respects the two models look identical. However, if one wants to express the physics of the GLSM associated with the torsional orbifold in terms of the superfield basis of the non--torsional GLSM, one has to perform anomalous superfield redefinitions. Since, these superfield redefinitions have to be well defined in each patch where they are performed, the expression of the anomaly is harmless within the smooth resolution phases. But in the orbifold phase this anomaly turns out not to be invariant under residual discrete $\Intr_2\times\Intr_2$ gauge transformations, precisely reproducing the torsion phases of the orbifold theory.

The story for the compact case is more involved. GLSMs for resolutions of the $T^6/\Intr_2\times\Intr_2$ orbifold have not explicitly appeared in the literature. Moreover, GLSMs for other compact orbifold resolutions have only been studied in the $(2,2)$ context. Therefore, before the question about discrete torsion on compact orbifold resolutions could be addressed, first resolution GLSMs for $T^6/\Intr_2\times\Intr_2$ had to be constructed. Contrary to the existing literature on compact orbifold resolutions, this was done immediately in the $(0,2)$ language. The simplest version of such a GLSM involves six gaugings on the worldsheet: three to define modified Weierstrass models to describe the underlying two--tori of the $T^6$ and three exceptional gaugings associated with the blowup process. In order to make comparisons with the non--compact situations most transparant, the same blowup modes were chosen as in the non--compact case, {\em i.e.}\ non--oscillatory twisted states. To pass all consistency conditions this resulted in a more complicated bundle that shares both features of line bundles on the resolved fixed two--tori as well as the standard embedding on the underlying two--tori of the $T^6$.

The resolution GLSM of the $T^6/\Intr_2\times\Intr_2$ with discrete torsion was obtained in a similar fashion as its non--compact analog: the exceptional gauge charges were flipped, while the other three gauge charges remained unchanged. As a consequence the resolution GLSM associated with the torsional orbifold suffers from mixed gauge anomalies. These anomalies can be cancelled by superfield dependent FI--terms in the GLSM globally. This signifies that the target space geometry has torsion in the sense that the three--form $H$--flux is non--zero. Moreover, given the GLSM chiral superfield content, the field dependent FI--terms need to involve logs of chiral superfields. As argued in the past, this signals that there are NS5--branes in the system. The structure of these logs can be taken such that these NS5--branes are located at the resolved exceptional cycles. In the orbifold limit they would disappear inside these singularities.

It is striking to see the differences of the effect of discrete torsion in the resolution process for non--compact and compact orbifolds. In the non--compact case apart from a physically irrelevant flip of charge conjugated states the non--torsional and torsional orbifold resolutions are to a large extent indistinguishable: only the relative signs of the gauge charges of the chiral and chiral Fermi multiplets distinguish them. In the compact case the GLSM associated to the torsional orbifold is really physically different from the non--torsional one as the mixed gauge anomalies and the related NS5--branes signify. These differences may be explained by realising that in the non--compact case the effect of flux can be pushed off to infinity while in the compact case this is impossible.

\subsubsection*{Outlook}

The work presented here can be extended in a number of ways. 

First of all, it would be interesting if it is possible by other means to show that the emerged picture that NS5--branes are located at the resolved singularities of the resolved torsional orbifold can be corroborate. And it would be interesting to confirm the interpretation of the coefficients that allow to shift mixed gauge anomalies around as moving around the NS5--branes of the resolved geometry. In addition, it would be interesting to investigate what the geometrical consequences are of the back reaction induced by the log--dependent FI--terms.  

In this paper the focus was on only very simple bundles (line bundles combined with bundles that are on the (2,2) locus, hence closely related to the standard embedding). However, the procedures used here could be applied to other gauge backgrounds as well. In particular, by choosing other blowup modes, for example, those with oscillator excitations, see {\em e.g.}~\cite{GrootNibbelink:2010qut}. 

Moreover, in this work only the discrete torsion between two orbifold twists was considered. For possible applications of the spinor--vector duality on smooth geometries other generalised discrete torsion phases would be of interest. First attempts in this direction were performed using effective field theory techniques in~\cite{Faraggi:2021fdr}. Such phases are between orbifold twists and torus translations and associated Wilson lines or among two different torus translations. This requires that within the GLSM distinctions between the various (resolved) fixed tori can be made. Clearly, this is possible in the maximal full resolution model, which treats all 48 (resolved) fixed tori independently, or in certain full resolution GLSM that have a certain number of additional gauging so that at least some fixed two--tori in certain directions can be distinguished. The effect of the Wilson lines would then be that the exceptional GLSM charges (dictated by the shifted twisted state momenta) are not the same at the different fixed tori. Then, just like in the models considered here, the effect of generalised discrete torsion is that particular states are projected out or in, leading to different charge assignments for the Fermi multiplets. Presumably, the consequences of these differences could then be analysed in much the same fashion as done in the current work.

\subsubsection*{Acknowledgements} 

SGN would like to thank the University of Liverpool for the kind hospitality during the completion of this work. 
We have benefitted from enlightening discussions with Sav Sethi during the String Phenomenology 2022 conference at Liverpool and subsequent email exchanges.  
We would also like to thank Eric Sharpe for useful questions and comments.

\appendix 
\addtocontents{toc}{\protect\setcounter{tocdepth}{1}}

\setcounter{equation}{0}
\section{Elements of $\boldsymbol{(0,2)}$ sigma models}
\label{sc:(0,2)models}

\subsection{$\boldsymbol{(0,2)}$ superspace} 

The $(0,2)$ superspace is spanned by a complex fermionic variable $\gth^+$ and its conjugate $\bgth^+$ of positive chiralilty in two dimensions and worldsheet coordinates $\gs = \sfrac 1{\sqrt{2}}(\gs_1+ \gs_0)$ and $\bgs = \sfrac1{\sqrt{2}}(\gs_1 - \gs_0)$. Using their derivatives denoted by $\der_+, \bder_+, \der = \sfrac 1{\sqrt{2}}(\der_1 + \der_0)$ and $\bder = \sfrac1{\sqrt{2}}(\der_1-\der_0)$, respectively, super covariant derivates $D_+$ and $\bD_+ = - (D_+)^\dag$ can be defined as 
\equ{
D_+ = \der_+ - i \bgth^+\, \der~, 
\qquad 
\bD_+ = \bder_+ - i \gth^+\, \der~, 
\qquad
\big\{ \bD_+, D_+ \big\} = -2i\, \der~.  
}
These super covariant derivatives anti--commute with the supercharges 
\equ{
Q_+ = \der_+ + i \bgth^+\, \der~, 
\qquad 
\bQ_+ = \bder_+ + i \gth^+\, \der~. 
}
The supercharges generate the $(0,2)$ super algebra
\equ{
\big\{ \bQ_+, Q_+ \big\} = 2\, P~, 
}
where $P = i\, \der$ is the right moving momentum generator.

\subsection{$\boldsymbol{(0,2)}$ superfields}

A general $(0,2)$ superfield $G$ is a complex function of $(0,2)$ superspace on which supersymmetry act as 
\equ{
\gd_\ge G = (\ge^+ Q_+ + \bge^{\,+} \bQ_+ ) G~. 
}
Consequently sums, products and super covariant derivatives of superfields are again superfields.

The components of a superfield are defined by taking a number of super covariant derivates and then set all $\gth^+$ and $\bgth^+$ to zero which is denoted as $|_+$. A superfield $G$ is called bosonic (fermionic) if its lowest component $G|_+$ is bosonic (fermionic).

There are four fundamental multiplets of $(0,2)$ supersymmetry: the chiral multiplet, the chiral Fermi multiplet, the vector multiplet and the Fermi gauge multiplet:

\subsubsection*{Chiral multiplet}
 
A chiral multiplet $\gF$ and its conjugate $\bgF$ are bosonic superfields defined by the chirality constraints: 
\equ{
\bD_+ \gF = 0~, 
\qquad 
D_+ \bgF = 0~.   
}
Their components, 
\equ{
z = \gF|_+~, 
\quad 
\gf = \sfrac 1{\sqrt{2}}D_+ \gF |_+~, 
\qquad 
\bz = \bgF|_+~, 
\quad 
\bgf = -\sfrac 1{\sqrt{2}}D_+ \bgF |_+~, 
}
are a complex scalar $z$, a negative chiral (right--moving) complex spinor $\gf$ and their conjugates $\bz$ and $\bgf$.

\subsubsection*{Chiral Fermi multiplet} 

A chiral Fermi multiplet $\gL$ and its conjugate $\bgL$ are fermionic superfields defined by the chirality constraints: 
\equ{
\bD_+ \gL = 0~, 
\qquad 
D_+ \bgL = 0~.
}
Their components, 
\equ{
\gl = \gL|_+~, 
\quad 
h = \sfrac 1{\sqrt{2}}D_+ \gL |_+~, 
\qquad 
\bgl = -\bgL|_+~, 
\quad 
\bh = \sfrac 1{\sqrt{2}}\bD_+ \bgL |_+~, 
}
are a positive chiral (left--moving) complex spinor $\gl$, a complex scalar $h$ and their conjugates $\bgl$ and $\bh$.

\subsubsection*{Vector multiplet}

The vector multiplet $(V,A)$ consists of two real bosonic superfields $V$ and $A$ subject to a bosonic super gauge transformation
\equ{
V \ra V - \sfrac 12\big(\gTh + \bgTh\big)~, 
\qquad 
A \ra A + \sfrac i2 \bder \big(\gTh - \bgTh\big)~, 
}
with a chiral superfield $\gTh$ gauge parameter and its conjugate $\bgTh$. (Non--Abelian gauge superfields are not considered in this work.) Their components are 
\equ{
\gTh|_+ = \gth = \sfrac12\, a + i\, \ga~, 
\qquad 
\sfrac1{\sqrt{2}} D_+ \gTh|_+ = \gz~, 
\qquad 
\bgTh|_+ = \gth = \sfrac 12\, a - i\, \ga~, 
\qquad 
-\sfrac1{\sqrt{2}} D_+ \bgTh|_+ = \bgz~, 
}
where $a$ and $\ga$ are real fields.  
The two dimensional gauge field components are identified as 
\equ{
A_\gs = \sfrac 12 \big[\,\bD_+, D_+ \big] V |_+~, 
\quad
A_\bgs = A|_+~, 
}
which transform as 
\equ{
A_\gs \ra A_\gs - \der \ga~, 
\qquad 
A_\bgs \ra A_\bgs - \bder \ga~.
}
The super field strengths 
\equ{
F = - \sfrac 12 \bD_+ \big(A - i \bder V\big)~, 
\qquad 
\bF = \sfrac 12 D_+ \big(A + i \bder V\big)~, 
}
are super gauge invariant chiral Fermi multiplets, since by construction 
$\bD_+ F = D_+ \bF = 0$. Consequently, their components 
\equ{
F|_+ = \sfrac 1{\sqrt{2}} \gvf~, 
\quad 
\bF|_+ = \sfrac 1{\sqrt{2}} \bgvf~, 
\quad 
D_+F|_+ =  \sfrac 12\big(D + i\, F_{\gs\bgs}\big)~, 
\quad
\bD_+\bF|_+ = \sfrac 12\big(D - i\, F_{\gs\bgs}\big)~, 
}
are identical in any gauge. In particular, $D = \sfrac 12\, [\bD_+,D_+] A|_+ - \der\bder V|_+$ and $F_{\gs\bgs} = F_{01}$.

The super gauge transformation can be used to set some of the components of the vector multiplet to zero: $V|_+ = D_+V|_+ = \bD_+ V|_+ = 0$. In this so--called Wess--Zumino (WZ) gauge all quadratic and higher powers of $V$ vanish. Since $V$ is a real superfield the WZ gauge does not fix the super gauge transformations completely, there is a residual gauge transformation with $\gTh = i \ga$.

\subsubsection*{Fermi gauge multiplet}

A Fermi gauge multiplet $\gS$ and its conjugate $\bgS$ are complex fermionic superfields subject to fermionic super gauge transformations 
\equ{
\gS \ra \gS - \gX~, 
\qquad 
\bgS \ra \bgS - \bgX~,
}
where $\gX$ is a Fermi multiplet and $\bgX$ its conjugate. The associated super field strength $\gY$ and its conjugate 
\equ{
\gY =  \bD_+ \gS~, 
\qquad 
\bgY = D_+ \bgS~, 
}
are inert under the fermionic gauge transformations. Their components are 
\equ{
s = \sfrac 1{\sqrt{2}} \gY|_+~, 
\qquad 
\bs = \sfrac 1{\sqrt{2}} \bgY|_+~, 
\qquad
\gch = \sfrac 12 D_+ \gY|_+~, 
\qquad 
 \bgch = \sfrac 12 \bD_+ \bgY|_+~. 
}

Using the fermionic gauge transformations, the following components of the Fermi gauge multiplet $\gS$ are set to zero in the WZ--gauge: $\gS|_+ = D_+ \gS|_+ = 0$.

\subsection{Super conformal transformations and scaling dimensions}
\label{sc:SCtransformations} 

Real conformal transformations of the worldsheet coordinates 
\equ{
\gs \ra f(\gs)~,
\qquad \bgs \ra \bff(\bgs)~, 
}
are characterized by two real functions $f(\gs)$ of $\gs$ only and $\bff(\bgs)$ of $\bgs$ only. Consequently, their differential and derivatives transform 
\equ{
\d\gs \ra  \go^{-1}\, \d\gs~
\qquad 
\d\bgs \ra \bgo^{-1}\, \d\bgs~, 
\qquad 
\der \ra \go\, \der~, 
\qquad 
\bder \ra \bgo\, \bder~, 
} 
where $\go = (\der f)^{-1}$ and $\bgo = (\bder \bff)^{-1}$. Moreover, since $\gth^+$ is a complex parameter, there is a phase transformation, often dubbed R--symmetry,
\equ{
\gth^+ \ra e^{i \gk}\, \gth^+~,
\qquad 
\bgth^+ \ra e^{-i \gk}\, \bgth^+~,
}
with $\gk \in \Real$. 
Requiring that the algebra of the super covariant derivatives transforms consistently with this implies: 
\equ{ 
D_+ \ra \go^{\sfrac 12}e^{-i \gk}\, D_+~, 
\qquad 
\bD_+ \ra \go^{\sfrac 12}e^{+i \gk}\, \bD_+~. 
}
The left-- and right--Weyl dimensions and the R--charge $(\cL,\cR,\text{R})$ (often collectively referred to as Weyl charges) of a general complex superfield $G$, defined as
\equ{
G \ra \bgo^{\cL}  \, \go^{\cR} \, e^{i \text{R}\gk} \, G 
}
identify how it responds to these conformal transformations. Real superfields are necessarily inert under the R--symmetry. The Weyl and R--charges of the superfields used in this work can be found in Table~\ref{tb:Superfields}.

\subsection{Scale invariant matter actions}

\subsubsection*{Scale invariant superspace integrals}

Any real bosonic superfield $R$ can be used to form a supersymmetric invariant by an integral over the full superspace: 
\equ{
S_\text{full\,superspace}  = \int \d^2 \gs \d^2 \gth^+ \, R = \int \d^2 \gs\, \bD_+ D_+ R |_+~. 
}
This action is gauge invariant if $R$ caries no gauge charges and scale invariant if it has Weyl and R--charges $(+1,0,0)$. 

Any chiral Fermi superfield $\gO$ can be used to form a supersymmetric invariant by an integral over the chiral superspace: 
\equ{
S_\text{chiral\,superspace} = \int \d^2\gs \d \gth^+\, \gO + \int \d^2 \gs \d \bgth^+\, \bgO = 
\int \d^2 \gs\, \big[ D_+ \gO + \bD_+ \bgO \,\big]|_+~. 
}
This is gauge invariant if $\gO$ carries no gauge charges and conformally invariant  if it has Weyl and R--charges $(+1,+\sfrac 12,+1)$.

\subsubsection*{Chiral superfield action}

The gauge interactions of chiral superfields $\gF^a$ and their conjugates $\bgF^\ua$ with Abelian vector multiplets $(V, A)_i$ are parameterized by the gauge charges $(q^i)_a$. In order to reduce the abundance of indices, interpret $q\!\cdot\! V$ as the diagonal matrix with on the diagonal $\sum_i (q^i)_a V_i$ and interpret $\gF$ as standing and $\bgF$ as lying vectors of $\mathrm{N}_\gF$ chiral superfields and their conjugates, respectively. Their super gauge transformations read  
\equ{
\bgF \ra \bgF\, e^{q\cdot \bgTh}~, 
\qquad 
\gF \ra e^{q\cdot \gTh}\, \gF~.
}
Their super gauge invariant kinetic action is given by 
\equ{ \label{eq:GLSM_chiral} 
S_\text{chiral} = \dfrac i4 \int\d^2\gs\d^2\gth^+\, \Big[ 
\bgF \, e^{2q\cdot V} \bcD \gF 
- \bcD\,\bgF\, e^{2q\cdot V} \gF 
\Big]~,  
}
in terms of the super gauge covariant derivatives of the chiral superfields and their conjugates 
\equ{ 
\bcD \gF = \bder\gF +  q\!\cdot\! \big(\bder V+i A\big) \gF~, 
\qquad 
\bcD \bgF = \bder\,\bgF +  \bgF \,q\!\cdot\! \big(\bder V-i A\big)~. 
}

\subsubsection*{Chiral Fermi superfield action}

The gauge and Fermi gauge interactions of the chiral Fermi superfields $\gL^m$ and their conjugates $\bgL^\um$ with the Fermi gauge multiplets are parameterised by the gauge charges $(Q^i)_m$ and holomorphic functions $U^{mI}(\gF)$. The super gauge and super fermionic gauge transformations of them read  
\equ{ \label{eq:GaugeTransFermiMultiplet} 
\bgL \ra \big(\bgL + \bgX \!\cdot\! \bU(\bgF) \big)\, e^{Q\cdot \bgTh}~, 
\qquad 
\gL \ra e^{Q\cdot \gTh}\, \big(\gL + U(\gF) \!\cdot\! \gX\big)~. 
}
Here the notation $\big( U(\gF)\!\cdot\!\gX)^m = U^{mI}(\gF)\, \gX_I$ is employed. 
The gauge charges of the holomorphic functions $U^{mI}(\gF)$ are $(Q^I)_m$ as well. Their super gauge invariant kinetic action is given by 
\equ{ 
S_\text{Fermi} = -\dfrac 12 \int\d^2\gs\d^2\gth^+\, 
\big(\bgL+ \bgS \!\cdot\! \bU(\bgF) \big)\, e^{2Q\cdot V}  \big(\gL + U(\gF) \!\cdot\! \gS\big)~. 
}

\subsubsection*{FI actions}

The Fayet--Ililopoulos (FI) action is given by the chiral superspace integral 
\equ{ \label{eq:GLSM_FI} 
S_\text{FI} = 
%
 \int\d^2\gs\d\gth^+\, W_\text{FI} + \text{c.c.}~, 
\quad 
W_\text{FI} = \gr(\gF)\!\cdot\! F  + (\gk(\gF)\!\cdot\!\gY) \gL ~, 
}
where $(\gk(\gF)\!\cdot\!\gY) \gL = \gk_{Im}(\gF)\gY_I \gL^m$ employing holomorphic functions $\gr_i(\gF)$ and $\gk_{Im}(\gF)$ of the chiral superfields $\gF^a$. The lowest components of $\gr(\gF)$, 
\equ{
\gr_i|_+ = \sfrac 12\, r_i + i\, \gb_i~, 
\qquad 
\bgr_i|_+ = \sfrac 12\, r_i - i\, \gb_i~, 
}
couple to the auxiliary field $D^i$ and the gauge field strength $F^i_{01}$, respectively: 
\equ{ 
\int\d\gth^+\, \gr(\gF)\!\cdot\! F + \int \d\bgth^+\, \bgr(\bgF)\!\cdot\! \bF \supset  
\sfrac 12\, r \!\cdot\! D - \gb \!\cdot\! F_{01}~, 
}
where $\supset$ indicates that the expression on the left includes terms given on the right.

Only when $\gr_i(\gF)$ are super gauge invariant and $\gk_{IA}(\gF)$ carry the opposite charges as $\gL^a$, the FI action is gauge invariant. This action is only invariant under fermionic gauge transformation if 
\equ{
\gk_{Im}(\gF) U^{mJ} = 0~, 
}
for all $I, J$. A worldsheet variant of the Green--Schwarz mechanism involves chiral superfield functions $\gr_i(\gF)$ that transforms as shifts under super gauge transformations.

\subsection{None scale invariant actions}

In GLSMs also actions are used that are not scale invariant. They involve parameters of mass dimension one or two in two dimensions. For simplicity all these parameters are assumed to be equal to $m$ or $|m|^2$, depending on whether these action are chiral or full superspace integrals. Consequently, conformal invariance is broken by these actions unless these parameters are send to either $0$ or $\infty$. Here, the conformal limit is taken to be the strong coupling limit $|m|\ra\infty$. In a more precise analysis one should study the renormalisation of the theory to understand if a conformal limit exists~\cite{Silverstein:1994ih,Silverstein:1995re}.

\subsubsection*{Gauge multiplet actions}

Abelian vector multiplets $(V,A)_i$ have kinetic actions 
\equ{
S_\text{gauge} =   \frac 1{2|m|^2}\, \int\d^2\gs\d^2\gth^+\, \bF F~. 
}
The kinetic terms for Fermi gauge multiplets $\gS_I$ are given by 
\equ{
S_\text{Fermi\,gauge} =  \frac 1{2|m|^2} \, \int\d^2\gs \d^2\gth^+\, \bgY \bder \gY~. 
}

\subsubsection*{Superpotentials}

To introduce gauge invariant superpotential actions, chiral superfields $\gPs^A$ and Fermi superfields $\gG^M$ are needed. They are given in Table~\ref{tb:Superfields}. The super gauge transformations of $\gPs$ read  
\equ{
\bgPs \ra \bgPs\, e^{\textsf{q}\cdot \bgTh}~, 
\qquad 
\gPs \ra e^{\textsf{q}\cdot \gTh}\, \gPs~.
}
The super gauge and super fermionic gauge transformations of $\gG$ are given by  
\equ{
\bgG \ra \big(\bgG + (\bgX \!\cdot\! \bW(\bgF))\,\bgPs \big)\, e^{\textsf{Q}\cdot \bgTh}~, 
\qquad 
\gG \ra e^{\textsf{Q}\cdot \gTh}\, \big(\gG + (\gPs\,W(\gF) \!\cdot\! \gX)\big)~. 
}
Here $\big[\gPs\, (W(\gF)\!\cdot\! \gX)\big]^M = \gPs^A W_{A}{}^{IM}(\gF) \gX_I$ is parameterised by chiral superfield functions $W_{AI}{}^M(\gF)$.

The superpotential action contains two pieces associated to the target space geometry and the gauge bundle that supports it:  
\equ{ 
S_{SP} = 
%
%
m\, \int\d^2\gs\d\gth^+\, \Big( P_\text{geom} + P_\text{bundle} \Big) 
+ \text{c.c.}~, 
\quad 
P_\text{geom} =  \gG\, P(\gF)~, 
\quad 
P_\text{bundle} = \gPs\, M(\gF)\, \gL~. 
}
Here, $\gG$ and $\gPs$ are interpreted as lying vectors of Fermi multiplets $\gG^M$ and chiral multiplets $\gPs^A$, respectively; $P(\gF)$ as a standing vector of chiral superfield functions $P_M(\gF)$ and $M(\gF)$ as a matrix of chiral superfield functions $M_{Am}(\gF)$. This is gauge invariant if the functions $P_M(\gF)$ carry the opposite gauge charges as $\gG^M$ and $M_{Am}(\gF)$ the opposite gauge charges as $\gPs^A \gL^m$. The superpotential action is only invariant under fermionic gauge transformations if~\eqref{eq:SPfermigaugeInv} holds.

The structure of the superpotential is dictated by a large extend by the Weyl charges: The R--charge implies that $\gPs^A$ and $\gG^M$ can only appear linearly in this expression. However, the superpotential is not conformal invariant, hence the mass parameter $m$ sits out front. This implies that in the conformal limit the superpotential has to vanish strictly.

 To complete the description also kinetic terms need to be added for the field $\gPs$ and $\gG$. The super gauge invariant kinetic action for $\gPs$ is given by 
\equ{ 
S_\text{chiral} = \dfrac i{4} \int\d^2\gs\d^2\gth^+\, \Big[ 
\bgPs \, e^{2\textsf{q}\cdot V} \bcD \gPs 
- \bcD\,\bgPs\, e^{2\textsf{q}\cdot V} \gPs 
\Big]~.  
}
The super gauge invariant kinetic action for $\gG$ is given by 
\equ{ 
S_\text{Fermi} = -\dfrac 1{2} \int\d^2\gs\d^2\gth^+\, 
\big(\bgG+ \bgS \!\cdot\! \bW(\bgF) \bgPs \big)\, e^{2\textsf{Q}\cdot V}  \big(\gG + \gPs W(\gF) \!\cdot\! \gS\big)~. 
}
The are both scale invariant.

\subsection{(0,2) non--linear sigma models}
\label{sc:FromGtoNLSM} 

The general action of a $(0,2)$ non--linear sigma model consists of two parts: an action for the chiral superfields $\BgF^\ga$\,, $\ga = 0,\ldots, 3$, and Fermi multiplets $\BgL^\gm$\,, $\gm = 1,\ldots, 16$. Here the scalar components of the chiral multiplets are interpreted as the local coordinates of the target space manifold $\cM$ and the fermionic components of the Fermi multiplets as the local coordinates in a section of the bundle $\cV$ in the same coordinate patch.

\subsubsection*{Torsional non--linear sigma models}

The most general conformal $(0,2)$ action of the chiral multiplets 
\equ{ \label{eq:NLSM_chiral} 
S_\text{n.l.\,chiral} = \frac i4 \int\d^2\gs\d^2\gth^+\, 
\Big[ 
K(\BgF,\BbgF)\, \bder \BgF - \bder \BbgF\, \bK(\BgF,\BbgF) 
\Big]~, 
}
are parameterised in terms of a lying complex vector function $K(\BgF, \BbgF)$ with entries $K_\ga(\BgF,\BbgF)$ and its conjugate, a standing vector $\bK(\BgF,\BbgF)$ with entries $\bK_\uga(\BgF,\BbgF)$. These functions are defined modulo additions  
\equ{
\bK(\BgF, \BbgF) \ra \bK(\BgF, \BbgF) + \bk(\BbgF)~, 
\qquad 
K(\BgF, \BbgF) \ra K(\BgF, \BbgF) + k(\BgF)~
}
of holomorphic vector functions $k(\BgF)$ and $\bk(\BbgF)$, as this would modify the full superspace integrant by a sum of a chiral superfield and its conjugate which vanishes. The superfield functions $K(\BgF,\BbgF)$ and $\bK(\BgF,\BbgF)$ can be thought of as prepotentials for the metric 
\equ{
G_{\uga\ga} = 
\sfrac 12 \Big( \bK_{\uga,\ga} + K_{\ga,\uga} \Big)
}
and the Kalb--Ramond two--form $B_2$
\equ{
B_{\uga\ga} = 
\sfrac 12 \Big( \bK_{\uga,\ga} - K_{\ga,\uga} \Big)~, 
\qquad 
B_{\ga\gb} = 
\sfrac 12 \Big( K_{\ga,\gb} - K_{\gb,\ga} \Big)~, 
\qquad 
B_{\uga\ugb} = 
\sfrac 12 \Big( \bK_{\uga,\ugb} - \bK_{\gb,\uga} \Big)~, 
}
combined, as can be seen by working out the kinetic action for the scalar components of the chiral superfields. The representation of the action for the scalar components is not unique due to $B_2$-field gauge transformations. A gauge can be chosen such that the components of the $B_2$--field with purely (anti--)holomorphic indices are absent. The non--vanishing components of the gauge invariant three--form field strength $H_3 = \d B_2$ can also be expressed in terms of these prepotential functions: 
\equ{
H_{\ga\gb\ugg} = H_{\gb\ugg\ga} = H_{\ugg\ga\gb} = 
K_{\ga,\gb\ugg} - K_{\gb,\ga\ugg}~, 
\qquad 
H_{\uga\ugb\gg} = H_{\ugb\gg\uga} = H_{\gg\uga\ugb} = 
\bK_{\uga,\ugb\gg} - \bK_{\ugb,\uga\gg}~. 
}
if some of these components are non--zero the manifold possesses torsion.

\subsubsection*{Chiral superfield interactions with Fermi multiplets}

The most general Weyl invariant action of Fermi multiplets is given by
\equ{
S_\text{n.l.\,Fermi} = 
- \frac 12 \int\d^2\gs \d^2\gth^+\, 
\Big\{
\BbgL\, N(\BgF,\BbgF)\, \BgL + \frac 12\, \BgL^T\, n(\BgF,\BbgF)\, \BgL + \frac 12\, \BbgL\, \bn(\BgF, \BbgF)\, \BbgL^T~,  
\Big\}
}
parameterised by an Hermitean matrix $N(\BgF,\BbgF)$ with entries $N_{\ugm\gn}(\BgF,\BbgF)$ assumed to be invertible and a complex anti--symmetric matrix $n(\BgF,\BbgF)$ with holomorphic indices $n_{\gm\gn}(\BgF,\BbgF)$ and its conjugate $\bn(\BgF,\BbgF)$ with entries $\bn_{\ugm\ugn}(\BgF,\BbgF)$. They can be thought of as the prepotentials for the target space gauge fields 
\equ{
A_\ga(N) = N^{-1} N_{,\ga}~, 
\qquad 
A_\uga(N) = N^{-1} N_{,\uga}~, 
\qquad
A_\uga(n) = n_{,\uga}
\qquad 
A_\ga(\bn) = \bn_{,\ga}~. 
}

\subsubsection*{From (0,2) GLSMs to (0,2) NLSMs}

By integrating out the gauge superfields $(0,2)$ GLSMs can be related to $(0,2)$ NLSMs. In particular, the equations of motion of $A$ lead to the constraints~\eqref{eq:ConstraintV} in the conformal limit. Then, by applying partial integrations on the derivative $\bder$ in the remaining ($A$ independent) terms in the FI--interaction~\eqref{eq:GLSM_FI} and combining them with the remaining kinetic terms of the chiral multiplets~\eqref{eq:GLSM_chiral}, these actions can be cast in the form of the NLSM action~\eqref{eq:NLSM_chiral} with the prepotentials 
\equ{ 
K_a = \big( \bgF\, e^{2q\cdot V}\big)_a +  2\,\gr_{,a}\!\cdot\! V~, 
\qquad 
\bK_\ua = \big( e^{2q\cdot V} \gF\big)_\ua + 2\, \bgr_{,\ua}\!\cdot\! V~. 
}
To see if these prepotentials for the metric and the B--field possess torsion, we compute the anti--symmetrised derivative 
\equ{
K_{[a,b]} = K_{a,b} - K_{b,a} = 
\big( \bgF\, e^{2q\cdot V} q\!\cdot\!V_{,[b}\big){}_{a]} +  2\,\gr_{,[a}\!\cdot\! V_{,b]}~. 
}
This expression can be simplified by taking the partial derivative w.r.t.\ $\gF^a$ of equation~\eqref{eq:ConstraintV} and after that contracting it with $V_{i,b}$. This gives
\equ{
\big( \bgF\, e^{2q\cdot V} q\!\cdot\!V_{,b}\big){}_a + 
2\bgF\, e^{2q\cdot V} (q\!\cdot\!V_{,a}) (q\!\cdot\!V_{,b}) \gF =  \gr_{,a} \!\cdot\! V_{,b}~, 
}
hence anti--symmetrised: 
\equ{ 
\big( \bgF\, e^{q\cdot V} q\!\cdot\!V_{,[b}\big){}_{a]} = \gr_{,[a}\!\cdot\! V_{,b]}~.
}
From which in general it may be concluded, that there will be torsion if the FI--functions $\gr_i(\gF)$ are not constant 
\equ{
K_{[a,b]} =  3\, \gr_{,[a}\!\cdot\! V_{,b]}~. 
}
From this the three--form $H$ expression~\eqref{eq:Hflux} follows immediately.

\setcounter{equation}{0}
\section{Anomalies in two dimensional GLSMs}
\label{sc:Anomalies}

\subsection{Chiral anomaly} 

Let $\gps$ be a Dirac fermion in two dimensions and $\bgps$ its conjugate. Consider the chiral transformation 
\equ{ 
\gps \ra e^{i\ga\, \frac{\Id+\tgg}2}\, \gps~, 
\qquad 
\bgps \ra \bgps\, e^{-i\ga\, \frac{\Id+\tgg}2}~. 
}
Here $\tgg = \gg^0\gg^1$ is the chirality operator in two dimensions satisfying $\tgg^2=\Id$. The anti--symmetrised product product of two gamma matrices is proportion to this operator:
\equ{ 
\gg^{\gm\gn} = \sfrac 12\, [\gg^\gm, \gg^\gn] = \ge^{\gm\gn}\, \tgg~, 
}
where $\ge^{\gm\gn} = -\ge^{\gn\gm}$ is the anti--symmetric epsilon tensor in two with the normalisation $\ge^{01} =1$. The Dirac operator of this fermion is assumed to couple chirally to a gauge field $A_\gm$: 
\equ{
D\Slashed = \partial\Slashed + i\, A\slashed\, \frac{\Id+\tgg}2~, 
}
where $A\Slashed = \gg^\gm A_\gm$ as usual. Note that 
\equ{ 
D\Slashed^2 = D^2 + \frac i2\, \tgg\, \ge^{\gm\gn} F_{\gm\gn}~,
}
where $i F_{\gm\gn} = [D_\gm,D_\gn]$ is the invariant gauge field strength or expressed as a two form
\equ{
F_2 = \sfrac 12\, F_{\gm\gn} \d \gs^\gm \d\gs^\gn = \sfrac 12\, \ge^{\gm\gn}F_{\gm\gn}\, \d^2\gs = F_{01}\, \d^2\gs~. 
}

If the path integral measure 
\equ{ 
\cD\gps\cD\bgps \ra \cD\gps\cD\bgps\, e^{i \cA_\text{chiral}}
}
is not invariant under this transformation, the chiral transformation is said to be anomalous. The anomaly can be expressed as the trace 
\equ{
 \cA_\text{chiral} = \Tr[  \ga\, \tgg] 
}
over both the full Hilbert and spinor space. This trace needs to be regularised. In case of anomalies a standard procedure is to use Fujikawa's regularisation
\equ{ 
\cA_\text{chiral} = \int \d^2\gs\, \ga\, \tr \langle x| \tgg\, e^{D\slashed^2/M^2} |x\rangle~, 
}
where $M$ is a regulator mass taken to be infinitely large. Using a plane wave expansion with a momentum variable $p$, scaling it as $p \ra M\, p$ and keeping only the leading terms this expression can be evaluated to 
\equ{
\cA_\text{chiral} = \int \d^2\gs\, \ga \int \frac{\d^2p}{(2\pi)^2} e^{-p^2} 
\tr \Big[ \tgg\, \sfrac i2\, \tgg\, \ge^{\gm\gn} F_{\gm\gn} 
\Big] 
}
where all the $M$ dependence dropped out (after taking the limit $M\ra\infty$). Using the Gaussian integral 
\equ{
\int \d^2p\, e^{-p^2} = \gp~, 
}
the chiral anomaly can be expressed as
\equ{
\cA_\text{chiral} = \int \sfrac i2\, \ga\, \frac{F_2}{2\pi}~. 
}

\subsection{Super gauge anomalies}

The result for the chiral anomaly above can be used for chiral gauge theories as well where then the parameter $\ga$ is interpreted as the gauge parameter of a $U(1)$ symmetry. For left--moving charged fermion the result can immediately be taken over, while for a right--moving fermion the expression will have an additional minus sign. If we have a set of left-- and right--moving fermions with charges $Q_i$ and $q_i$ under a number of $U(1)$ gauge symmetries, the result generalises to 
\equ{
\cA_\text{gauge} = 
\int \sfrac i2\, \ga^i\,\mathcal{A}_{ij} \, \frac{F^j_2}{2\pi}~, 
}
where the anomaly matrix is given by 
\equ{ 
  \mathcal{A}_{ij} =Q_{i}\cdot  Q_{j}-q_{i}\cdot q_{j}~. 
}
Here the dot product indicates the sum over all charged left and right fermions present in the theory. Assuming the existence of a supersymmetric regulator, the general form of super gauge anomalies in two dimensions can be written as 
\equ{ \label{eq:SuperGaugeAnom} 
S_\text{anom} = 
\int\d^2\gs\d\gth^+\, \frac 1{4\gp}  \sum_{i,j} \mathcal{A}_{ij} \,\gTh^{i} F^{j}
+  \int\d^2\gs\d\bgth^+\,\frac 1{4\gp} \sum_{i, j}  \mathcal{A}_{ij}\, \bgTh^{i} \bF^{j}~. 
}

\setcounter{equation}{0}
\section{Charge matrices}
\label{sc:ChargeMatrices}

In Section~\ref{sc:C3Z22glsms} a number of so--called charge matrices are used to perform certain computations. In a given patch of a given phase of the GLSM a number of charged superfields are necessarily non--zero. Their charge matrices are given by:  
\sequ{eq:ChargeMatricesC3Z22}{
\cQ_{(O)} = \pmtrx{
~\sm1~ & ~0~ & ~0~ \\ 
 0 & \sm1 & 0 \\
 0 & 0 & \sm 1  
}~, 
\qquad 
\cQ_{(S)} = \pmtrx{
~0~ & ~\sfrac 12~ & ~\sfrac 12~ \\ 
\sfrac 12 & 0 & \sfrac 12 \\
\sfrac 12 & \sfrac 12 & 0 
}~, 
\\[2ex] 
\cQ_{(11)} = \pmtrx{
\,\sm1~ & ~0~ & ~0~ \\ 
\sfrac 12 & 0 & \sfrac 12 \\
\sfrac 12 & \sfrac 12 & 0 
}~, 
\quad 
\cQ_{(12)} = \pmtrx{
~\sm1~ & ~0~ & ~0~ \\ 
0 & \sfrac 12 & \sfrac 12 \\
\sfrac 12 & \sfrac 12 & 0 
}~, 
\quad
\cQ_{(13)} = \pmtrx{
~\sm1~ & ~0~ & ~0~ \\ 
0 & \sfrac 12 & \sfrac 12 \\
\sfrac 12 & 0 & \sfrac 12 
}~,
\\[2ex]
\cQ_{(21)} = \pmtrx{
\sfrac 12 & 0 & \sfrac 12 \\
~0~ & ~\sm1~ & ~0~ \\ 
\sfrac 12 & \sfrac 12 & 0 
}~, 
\quad 
\cQ_{(22)} = \pmtrx{
0 & \sfrac 12 & \sfrac 12 \\
~0~ & ~\sm1~ & ~0~ \\ 
\sfrac 12 & \sfrac 12 & 0 
}~, 
\quad
\cQ_{(23)} = \pmtrx{
0 & \sfrac 12 & \sfrac 12 \\
~0~ & ~\sm1~ & ~0~ \\ 
\sfrac 12 & 0 & \sfrac 12 
}~,
\\[2ex] 
\cQ_{(31)} = \pmtrx{
\sfrac 12 & 0 & \sfrac 12 \\
\sfrac 12 & \sfrac 12 & 0 \\
~0~ & ~0~ & ~\sm1~ 
}~, 
\quad 
\cQ_{(32)} = \pmtrx{
0 & \sfrac 12 & \sfrac 12 \\
\sfrac 12 & \sfrac 12 & 0 \\
~0~ & ~0~ & ~\sm1~ 
}~, 
\quad
\cQ_{(33)} = \pmtrx{
0 & \sfrac 12 & \sfrac 12 \\
\sfrac 12 & 0 & \sfrac 12 \\ 
~0~ & ~0~ & ~\sm1~ 
}~.
}
Their transposed inverse are: 
\sequ{eq:ChargeMatricesC3Z22TransposedInverse}{
\cQ_{(O)}^{-T} = \pmtrx{
~\sm1~ & ~0~ & ~0~ \\ 
 0 & \sm1 & 0 \\
 0 & 0 & \sm 1  
}~, 
\qquad 
\cQ_{(S)}^{-T}  = 
\pmtrx{
~\sm1~ & ~1~ & ~1~ \\ 
1 & \sm1 & 1 \\ 
1 & 1 & \sm1 
 }~, 
\\[2ex] 
\cQ_{(11)}^{-T} = \pmtrx{
\,\sm1~ & ~1~ & ~1~ \\ 
0 & 0 & 2 \\
0 & 2 & 0 
}~, 
\quad 
\cQ_{(12)}^{-T} = \pmtrx{
\,\sm1~ & ~1~ & \,\sm1~ \\ 
0 & 0 & 2 \\
0 & 2 & \sm2
}~, 
\quad
\cQ_{(13)}^{-T} = \pmtrx{
\,\sm1~ & \,\sm1~ & ~1~ \\ 
0 & 2 & 0 \\
0 & \sm2 & 2 
}~,
\\[2ex]
\cQ_{(21)}^{-T} = \pmtrx{
0 & 0 & 2 \\
~1~ & \,\sm1~ & \,\sm1~ \\ 
2 & 0 & \sm2
}~, 
\quad 
\cQ_{(22)}^{-T} = \pmtrx{
0 & 0 & 2 \\
~1~ & \,\sm1~ & ~1~ \\ 
2 & 0 & 0 
}~, 
\quad
\cQ_{(23)}^{-T} = \pmtrx{
\sm2 & 0 & 2 \\
\,\sm1~ & \,\sm1~ & ~1~ \\ 
2 & 0 & 0 
}~,
\\[2ex] 
\cQ_{(31)}^{-T} = \pmtrx{
2 & \sm2 & 0 \\
0 & 2 & 0 \\
~1~ & \,\sm1~ & \,\sm1~ 
}~, 
\quad 
\cQ_{(32)}^{-T} = \pmtrx{
\sm2 & 2 & 0 \\
2 & 0 & 0 \\
\,\sm1~ & ~1~ & \,\sm1~ 
}~, 
\quad
\cQ_{(33)}^{-T} = \pmtrx{
0 & 2 & 0 \\
2 & 0 & 0 \\ 
~1~ & ~1~ & \,\sm1~ 
}~.
}
The charge matrices associated to the superfields that define a given patch read:  
\sequ{eq:ChargeMatricesC3Z22conjugate}{
\widetilde{\cQ}_{(O)} = \pmtrx{
~0~ & ~\sfrac 12~ & ~\sfrac 12~ \\ 
\sfrac 12 & 0 & \sfrac 12 \\
\sfrac 12 & \sfrac 12 & 0 
}~, 
\qquad 
\widetilde{\cQ}_{(S)} = \pmtrx{
~\sm1~ & ~0~ & ~0~ \\ 
 0 & \sm1 & 0 \\
 0 & 0 & \sm 1  
}~, 
\\[2ex] 
\widetilde{\cQ}_{(11)} = \pmtrx{
0 & \sfrac 12 & \sfrac 12 \\
~0~ & ~\sm1~ & ~0~ \\ 
0 & 0 & \sm1
}~,
\quad 
\widetilde{\cQ}_{(12)} = \pmtrx{
\sfrac 12 & 0 & \sfrac 12 \\
~0~ & ~\sm1~ & ~0~ \\ 
0 & 0 & \sm1
}~, 
\quad 
\widetilde{\cQ}_{(13)} = \pmtrx{
\sfrac 12 & \sfrac 12 & 0 \\
~0~ & ~\sm1~ & ~0~ \\ 
0 & 0 & \sm1
}~, 
\\[2ex]
\widetilde{\cQ}_{(21)} = \pmtrx{
~\sm1~ & ~0~ &  ~0~ \\ 
0 & \sfrac 12 & \sfrac 12 \\
0 & 0 & \sm1
}~,
\quad 
\widetilde{\cQ}_{(22)} = \pmtrx{
~\sm1~ & ~0~ &  ~0~ \\ 
\sfrac 12 & 0 & \sfrac 12 \\
0 & 0 & \sm1
}~,
\quad 
\widetilde{\cQ}_{(22)} = \pmtrx{
~\sm1~ & ~0~ &  ~0~ \\ 
\sfrac 12 & \sfrac 12 & 0 \\
0 & 0 & \sm1
}~,
\\[2ex] 
\widetilde{\cQ}_{(31)} = \pmtrx{
~\sm1~ & ~0~ &  ~0~ \\ 
~0~ & ~\sm1~ & ~0~ \\ 
0 & \sfrac 12 & \sfrac 12 \\
}~,
\quad 
\widetilde{\cQ}_{(32)} = \pmtrx{
~\sm1~ & ~0~ &  ~0~ \\ 
~0~ & ~\sm1~ & ~0~ \\ 
\sfrac 12 & 0 & \sfrac 12 \\
}~,
\widetilde{\cQ}_{(33)} = \pmtrx{
~\sm1~ & ~0~ &  ~0~ \\ 
~0~ & ~\sm1~ & ~0~ \\ 
\sfrac 12 & \sfrac 12 & 0 \\
}~.
}

\bibliographystyle{paper}
{\small
\bibliography{paper}
}
\end{document}